\documentclass[aps,prd,onecolumn,superscriptaddress,preprintnumbers,nofootinbib, a4paper]{revtex4}

\usepackage[utf8x]{inputenc}

\usepackage{amsmath,bbm,latexsym,amssymb}
\usepackage{graphicx}

\usepackage{multirow} 

\usepackage{xparse}
\usepackage{xcolor}

\usepackage[hyperfootnotes=false]{hyperref}
\hypersetup{
    colorlinks=true,
    linkcolor=blue,
    filecolor=blue,      
    urlcolor=blue,
    citecolor=blue
}

\usepackage{listings}
\usepackage{booktabs}

\usepackage{physics}
\usepackage{bbold}

\newcommand{\be}{\begin{equation}}
\newcommand{\ee}{\end{equation}}
\newcommand{\ba}{\begin{eqnarray}}
\newcommand{\ea}{\end{eqnarray}}

\newcommand{\bs}{\begin{subequations}}
\newcommand{\es}{\end{subequations}}
\newcommand{\no}{\nonumber\\}

\newcommand{\bt}{\begin{table}[h!]}
\newcommand{\et}{\end{table}}

\newcommand{\bb}{\begin{bmatrix}}
\newcommand{\eb}{\end{bmatrix}}

\newcommand{\linkGIT}{\url{
https://github.com/andremilagre/BounDS.git}}

\newcommand{\AddrCFTP}{%
 Departamento de Física and CFTP, Instituto Superior Técnico\\
 Universidade de Lisboa, 
          Av. Rovisco Pais 1, 1049-001 Lisboa, Portugal }

\def\gsim{\raise0.3ex\hbox{$\;>$\kern-0.75em\raise-1.1ex\hbox{$\sim\;$}}}
\def\lsim{\raise0.3ex\hbox{$\;<$\kern-0.75em\raise-1.1ex\hbox{$\sim\;$}}}

\allowdisplaybreaks

\begin{document}

\preprint{CFTP/25-xxx}

\title{Perturbative unitarity for models with singlet and doublet scalars} 
\author{Carolina T. Lopes}\email{carolina.teixeira.lopes@tecnico.ulisboa.pt}
\affiliation{\AddrCFTP}
\author{André Milagre}\email{andre.milagre@tecnico.ulisboa.pt}
\affiliation{\AddrCFTP}
\author{João  P.~Silva} \email{jpsilva@cftp.ist.utl.pt}
\affiliation{\AddrCFTP}

\today

\begin{abstract}
We provide a complete description of perturbative unitarity bounds on the gauge-scalar sectors of
models with extra 
$SU(2)$
doublet, neutral singlet, and charged singlet scalars.
Such additions are very frequent in models beyond the Standard Model, and,
in particular, they are almost universal in models explaining the dark matter problem.
We propose a specific classification and minimal set of scattering matrices
containing all the relevant information.
We also developed a Mathematica implementation of our results,
{\tt BounDS},
and we use it to fully study a number of simple cases,
comparing with the literature, when available.
The Mathematica notebook {\tt BounDS}
is provided via a public 
\href{https://github.com/andremilagre/BounDS.git}{GitHub} 
repository.
\end{abstract}

\maketitle

\newpage

\section{\label{Intro}Introduction}

The Standard Model (SM) of particle physics has been extremely successful,
culminating in the discovery of a scalar particle consistent with the Higgs
boson~\cite{ATLAS:2012yve,CMS:2012qbp}.
Nevertheless, it is unlikely to be the final theory, and extensions
beyond the Standard Model are widely considered.
In particular,
the SM provides no explanation for dark matter \cite{ParticleDataGroup:2016lqr},
whose presence is inferred across a wide range of astronomical scales,
from kiloparsecs to the size of the observable Universe.
Key probes include the rotation curves of spiral galaxies,
the study of galaxy clusters,
and measurements of the Cosmic Microwave 
Background~\cite{Planck:2018vyg,Cirelli:2024ssz}. 

One of the most widely studied scenarios envisions dark matter as consisting
of new fundamental particles. The simplest scalar extension introduces a real,
neutral 
$SU(2)$
singlet with a global $\mathbb{Z}_2$ symmetry to ensure 
stability~\cite{Burgess:2000yq,GAMBIT:2017gge}.
Alternatively, a minimal fermionic option is a Weyl fermion, commonly referred
to as a sterile neutrino~\cite{Cirelli:2024ssz}.
Other proposed models include the Inert Doublet Model,
which extends the Higgs sector with a dark doublet~\cite{Ilnicka:2015jba};
multi-singlet SM extensions~\cite{Goncalves:2025snm};
the Next-to-Minimal Supersymmetric Standard Model~\cite{Ellwanger:2009dp};
and models featuring scalar or fermionic portals to hidden sectors~\cite{Arcadi:2024wwg}.
Most of these models include extra scalar fields,
doublet or singlet under the SM gauge group,
and often new discrete symmetries in order to stabilize the dark sector.
Any consistent exploration of these theories must begin with the enforcement 
of basic theoretical requirements, such as the boundedness from below of the
scalar potential, the existence of a global minimum, and perturbative unitarity,
the latter being the main focus of the present work.

In a consistent quantum field theory, probability conservation is encoded
in the unitarity of the $S$-matrix~\cite{Schwartz:2014sze,Logan:2022uus}.
In scalar extensions of the Standard Model, this requirement 
is particularly restrictive, as the presence of additional scalar degrees of
freedom can lead to scattering amplitudes that grow with energy.
Imposing perturbative unitarity, therefore, yields non-trivial constraints 
on the masses and couplings of the new scalars, 
ensuring the validity of perturbation theory.
For $2 \to 2$ scattering processes,
perturbative unitarity constraints can be 
applied to the partial-wave decomposition
of the amplitudes, 
with the strongest high-energy limits coming from the zeroth partial-wave
amplitude~\cite{Schwartz:2014sze,Logan:2022uus}.
In the high-energy limit, the equivalence theorem~\cite{Veltman:1989ud}
ensures that the scattering of longitudinally polarized gauge bosons can be
replaced by the corresponding Goldstone bosons, allowing these scalar amplitudes
to capture the dominant contributions to perturbative unitarity bounds.
Historically, Lee, Quigg, and Thacker applied this method to derive
an upper bound on the Higgs boson mass,
$M_H^2 \leq 8 \pi \sqrt{2}/(3 G_F) \approx 1\textrm{TeV}^2$~\cite{Lee:1977yc,Lee:1977eg}.
Perturbative unitarity has also been applied to the general two-Higgs-doublet model (2HDM)
with explicit CP violation~\cite{Ginzburg:2005dt,Kanemura:2015ska},
to the general $N$-Higgs-doublet model~\cite{Bento:2017eti},
also with fermions~\cite{Bento:2018fmy},
and to constrain large scalar multiplets \cite{Hally:2012pu,Milagre:2024wcg}.
Many other examples can be found in the literature,
including
\cite{LlewellynSmith:1973yud,Cornwall:1974km,Weldon:1984wt,Casalbuoni:1986hy,Casalbuoni:1987eg,Gunion:1990kf,Maalampi:1991fb,Kanemura:1993hm,Akeroyd:2000wc,Horejsi:2005da,ElKaffas:2007rq,Grinstein:2013fia,Nagai:2014cua,Englert:2016joy,Cacchio:2016qyh,Harris:2017ecz,Goodsell:2018tti,Goodsell:2018fex,Bento:2023weq,Urquia-Calderon:2024rzc,Blazek:2024udx}.
More recently, perturbative unitarity analyses have been 
extended beyond the traditional $2 \to 2$ limit.
For instance, Ref.~\cite{Bresciani:2025toe} studies unitarity bounds
in general $M \to N$ scattering processes with $M, N \geq 2$.
In particular, in inflationary models featuring a non-minimal 
coupling with the Ricci scalar, partial-wave unitarity bounds on 
$2 \to N$ processes have improved earlier estimates of the energy scale
at which new-physics effective interactions invalidate the 
perturbative expansion~\cite{Steingasser:2025txd}.

In this work, we extend the analysis of Ref.~\cite{Bento:2017eti} by
constructing an $SU(2)\times U(1)$-symmetric model with an arbitrary
number of scalar fields 
which may either be singlets or doublets under $SU(2)$. 
The hypercharge assignments are chosen such that the new scalars
are either electrically neutral or carry a single unit of electric charge.
With this framework, we then require all relevant $2\to2$ scattering
processes to satisfy perturbative unitarity at tree-level and in the
high-energy limit,
thereby deriving upper bounds on the quartic couplings of this class of theories.

In Section~\ref{sec:themodel}, we present the most general renormalizable
$SU(2)\times U(1)$-symmetric model in the high-energy limit,
composed of an arbitrary number of $SU(2)$ scalar doublets with
hypercharge $\tfrac{1}{2}$ and singlets that may either be electrically neutral,
or carry hypercharge 1.
We derive relations among the couplings by imposing Hermiticity
of the scalar potential and by eliminating redundant gauge-invariant operators.
In Section~\ref{sec:pwave}, we construct the scattering matrix from second derivatives
of the quartic potential, diagonalize it to obtain eigenamplitudes,
and impose perturbative unitarity, $|\Lambda|\leq 8\pi$, to constrain the quartic couplings.
We propose to classify states by the conserved quantum
numbers $|Q,Y,T\rangle$, corresponding to electric charge, hypercharge, and total isospin,
respectively.
In Appendix~\ref{app:REDUNDANT}, we illustrate the benefits of labeling states
not only by $Q$ and $Y$,
but also by total isospin $T$, using the Standard Model
as a case study in Appendix~\ref{app:SMexample}.
This classification facilitates the construction of scattering matrices and the
identification of independent channels, for which we present a minimal basis
(meaning that we discuss the selection of basis states that allow all independent eigenvalues to be determined).
We provide explicit formulas for the potential and the matrix elements of
all scattering processes allowed in our class of models. 
In Section~\ref{sec:Math}, we present the 
{\tt Mathematica} notebook {\tt BounDS} that allows for the
automatic calculation of the quartic part of the scalar potential 
\textit{and} of all
scattering matrices in models of this type;
the user must only input the particle content,
and, if needed, the transformation properties of the fields 
under additional flavour and/or generalized CP symmetries\footnote{
These can either be discrete or continuous, Abelian or non-Abelian.}.
We make the notebook publicly available at
\begin{center}
\linkGIT
\end{center}
In Section~\ref{particularcases}, we apply these methods to particular models
by specifying the scalar content and symmetries, and
we compare with results in the literature, when available.
We present our results in terms of the parameters of the potential.
In specific models,
it is sometimes possible to write the parameters of the potential in terms of scalar masses
and/or mixing angles.
That implies defining gauge-invariant terms of dimension two and/or three,
defining the vacuum of the theory,
writing all (neutral and charged) scalar mass matrices,
diagonalizing them to obtain masses and mixing angles in terms of potential parameters,
and then inverting those relations.
This must be done on a case-by-case basis and lies beyond the scope of this work.
We present our conclusions in Section~\ref{sec:conclusion}.

In Appendix~\ref{app:NOTEBOOK} we provide further details on our {\tt Mathematica} notebook.
Finally, Appendix~\ref{appD} addresses
the inclusion in the basis of additional quantum numbers,
when extra flavour symmetries,
such as $\mathbb{Z}_n$ or CP, are present.

\section{The Model}
\label{sec:themodel}

\subsection{Particle Content}
Consider a gauge theory symmetric under the $SU(2) \times U(1)$
group with an arbitrary number of scalar fields (and their charge conjugates)
that are either doublets or singlets of $SU(2)$.
We define the electric charge $Q$ as the sum of the third component of
isospin $T_3$, and the hypercharge $Y$, and restrict our discussion to models
where the scalar fields are either electrically neutral or carry a single unit of electric charge.
As a consequence, this theory may have $n_D$ $SU(2)$
scalar doublets with $Y=1/2$ which we denote by
\be
\Phi_i = \begin{pmatrix} \phi_i^+ , \, \phi_i^{0 }\end{pmatrix}^T, \quad i = 1, \ldots, n_D;
\ee 
$n_c$ $SU(2)$ complex scalar singlets with $Y=1$ which we denote by 
\be
\varphi_i^+,\quad i=1, \ldots, n_c;
\ee 
$n_n$ $SU(2)$ real scalar singlets with $Y=0$ which we denote by 
\be
\chi_i,\quad i=1, \ldots, n_n.
\ee 
Notice that models with $m$ $SU(2)$ complex scalar singlets with $Y=0$ are particular
cases of models with $n_n=2m$ real scalar singlets with 
additional $\mathbb{Z}_2$ symmetries;
hence, we do not treat them separately.
The oblique radiative corrections for this class of models has been thoroughly
studied in Refs.~\cite{Grimus:2007if,Grimus:2008nb}.

\subsection{Scalar potential}
In this work, we are interested in constraining the parameters of models 
like the one presented in Section~\ref{sec:themodel}
by imposing partial-wave unitarity bounds on $2\to 2$ scattering processes.
We perform all calculations in the high-energy regime, where the contributions
from propagators in the $s$, $t$, and $u$-channels vanish, deeming the quartic interactions in the potential the dominant contributions.
The most general renormalizable quartic part of the scalar potential of
such a theory can be written as:\footnote{For the simplest model
including the $\kappa_{ab,cd}$ terms, see Sec.~\ref{sec:2D1N1C} below.
}
\ba
V \supset V_4 &=&
\lambda_{ab,cd} \left(\Phi_a^\dagger \Phi_b\right)\left(\Phi_c^\dagger \Phi_d\right)+
\alpha_{ab,cd} \left(\varphi^-_a\varphi^+_b\right)\left( \varphi^-_c\varphi^+_d \right)+
\beta_{ab,cd}\left( \chi_a\chi_b \right)\left( \chi_c\chi_d \right)+
\no
&&
\delta_{ab,cd} \left(\Phi_a^\dagger \Phi_b\right)\left( \varphi^-_c\varphi^+_d \right)
+ \gamma_{ab,cd} \left(\Phi_a^\dagger \Phi_b\right)\left( \chi_c\chi_d \right) +
\zeta_{ab,cd} \left(\varphi^-_a\varphi^+_b\right)\left( \chi_c\chi_d \right) +
\no 
&& 
\kappa_{ab,cd} \left(\Phi_a^T \sigma_2 \Phi_b\right) \left(\varphi^-_c \chi_d\right) +
\kappa_{ab,cd}^* \left(\Phi_b^\dagger \sigma_2 \Phi_a^*\right) \left(\varphi^+_c \chi_d\right) ,
\label{eq:potential}
\ea
where $\sigma_2$ is the $2 \times 2 $ second Pauli matrix, 
and a sum over repeated indices is implied.

Although compact, not all couplings in this notation are independent.
By requiring the scalar potential to be Hermitian and 
rearranging gauge invariant bilinears\footnote{
By \textit{rearranging gauge invariant bilinears}, we mean that
\begin{equation}
	\left(\Phi_a^\dagger \Phi_b\right)\left(\Phi_c^\dagger \Phi_d\right)
	=
	\left(\Phi_c^\dagger \Phi_d\right)\left(\Phi_a^\dagger \Phi_b\right)
	\quad
	\Rightarrow
	\quad
	\lambda_{ab,cd} = \lambda_{cd,ab}.
\end{equation}
},
we derive the following relations between quartic couplings:
\ba
\label{propslambda}
\lambda_{ab,cd} &=& \lambda_{ba,dc}^*\ =\ \lambda_{cd,ab},\\
\label{propsalpha}
\alpha_{ab,cd} &=& \alpha_{ba,dc}^*\ =\ \alpha_{cd,ab}\ =\ \alpha_{ad,cb},\\
\label{propsbeta}
\beta_{ab,cd} &=& \beta_{(ab,cd)}^*\ =\ \beta_{(ab,cd)},\\
\label{propsdelta}
\delta_{ab,cd} &=& \delta_{ba,dc}^*\\
\label{propsgamma}
\gamma_{ab,cd} &=& \gamma_{ba,cd}^*\ =\ \gamma_{ab,dc},\\
\label{propszeta}
\zeta_{ab,cd} &=& \zeta_{ba, cd}^* \ =\ \zeta_{ab,dc},\\
\label{propskappa}
\kappa_{ab,cd} &=& -\kappa_{ba,cd},
\ea
where $(ab,cd)$ stands for any permutation of the indices.
In particular, these relations further imply
\begin{eqnarray}
\lambda_{aa,bb},\  \lambda_{ab,ba} &\in & \mathbb{R},\\
\alpha_{aa,bb} = \alpha_{ab,ba} &\in & \mathbb{R},\\
\beta_{ab,cd} &\in & \mathbb{R},\\
\delta_{aa,bb} &\in & \mathbb{R},\\
\gamma_{aa,cd} &\in & \mathbb{R},\\
\zeta_{aa,cd} &\in & \mathbb{R},\\
\kappa_{aa,cd} &=& 0.
\label{rrr}
\end{eqnarray}

\section{\label{sec:pwave}Partial wave unitarity bounds}

\subsection{Partial wave decomposition}
We consider $2 \to 2$ scattering processes between the scalars
of the theory defined in Section~\ref{sec:themodel}.
Let $A$, $B$, $C$, and $D$ be complex scalar fields
and let $a, b, c$, and $d$ be their corresponding flavour indices.
At tree-level, the amplitude for the process
\begin{equation}
	A_a \, B_b \to C_c \, D_d
\end{equation}
may have contributions from
$s$, $t$, and $u$-channels, and contact interactions.
However, as we take the limit where the Mandelstam variables 
$s$, $t$, and $u$ go to infinity, the contributions from 
$s$, $t$, and $u$-channels vanish, respectively~\footnote{
For an interesting example, see, for instance, Appendix A of \cite{MiguelMSc}.}.
Consequently, in the high-energy limit, only 
the quartic interactions involving the external scalars 
contribute to the tree-level amplitude, allowing us to identify
\begin{equation}
	\mathcal{M}\left[ A_a \, B_b \to C_c \, D_d \right]
	\ =\ 
	- \frac{\partial^4 V_4}
	{\partial A_a \,\partial B_b \,\partial C_c^* \, \partial D_d^*}.
\label{eq:Mabcd}
\end{equation}

Any square-integrable function may be expressed as 
an expansion in a complete set of basis functions.
To study perturbative unitarity,
it is particularly convenient to choose the basis 
of Legendre polynomials $P_J(\cos \theta)$, where $J$ is the 
total angular momentum of the final state,
and $\theta$ the scattering angle.
Expanding the amplitude in this basis defines the 
so-called partial-wave expansion~\cite{Hally:2012pu,Milagre:2024wcg}:
\begin{equation}
	\mathcal{M}(\cos\theta) = 
	16\pi
	\sum_{J=0}^\infty
	a_J \left(2J+1\right) P_J(\cos \theta).
\end{equation}
The numerical coefficients $a_J$ 
(also known as partial waves) 
are defined as
\begin{equation}
	a_J = \frac{2J+1}{32\pi} \int_{-1}^1
	\mathcal{M}(\cos\theta)\, P_J (\cos\theta) \ \mathrm{d}\cos\theta ,
\label{aJJJ}
\end{equation}
and are constrained by tree-level partial-wave unitarity 
through~\cite{Hally:2012pu,Milagre:2024wcg}
\begin{equation}
\left| a_J \right| \leq 1,\quad
0\leq \Im\left\{a_J\right\}\leq1,\quad
\textrm{and}\quad
\left| \Re\{a_J\} \right|\leq \frac{1}{2}.
\label{PWUB111}
\end{equation}

At tree-level, all $a_J$ are real, so the three inequalities in
Eq.~\eqref{PWUB111} reduce to $\left| a_J \right| \leq \frac{1}{2}$.
In the high-energy limit, the scattering amplitudes are independent
of the scattering angle, and since $P_0(\cos\theta)=1$,
the strongest bound arises from the zeroth partial wave, $a_0$.
The perturbative unitarity condition can then be written at the level of the amplitude as
\begin{equation}
	\left| a_0 \right| \ =\ 
	\frac{1}{16\pi} \Big|  \mathcal{M}\left[ A_a \, B_b \to C_c \, D_d \right]\Big|
	\ \leq \ \frac{1}{2},
	\label{PWUB}
\end{equation}
or, using Eq.~\eqref{eq:Mabcd}, at the level of the quartic couplings as
\begin{equation}
16\pi \left| a_0 \right| \ =\  
\left| N_{ab} N_{cd} \,
\frac{\partial^4 V_4}
{\partial A_a \,\partial B_b \,\partial C_c^* \, \partial D_d^*}
\right|\leq 8 \pi,
\label{UNI}
\end{equation}
where
\begin{equation}
	N_{ij} \equiv \frac{1}{\sqrt{2^{\delta_{ij}}}}
\end{equation}
is a symmetry factor that accounts for identical particles
either in the initial or final state.

\subsection{Scattering matrices}
\label{subsec:ScateringMatrices}

One may go a step further by employing the 
method of coupled-channel 
analysis \cite{Logan:2022uus,Ginzburg:2005dt,Kanemura:2015ska}.
This approach takes advantage of the fact 
that partial-wave unitarity
bounds may be imposed on any specific process, but also
to any superposition of states, provided they have 
the same quantum numbers.
By organizing the zeroth partial waves of such scatterings 
into a coupled-channel matrix, 
the perturbative unitarity condition translates into a bound on its eigenvalues.
The most stringent constraint is obtained by 
requiring that the modulus of the largest eigenvalue
to remain below $8 \pi$, as derived in Eq.~\eqref{UNI}.
Once again, note that $J=0$ partial waves
should be multiplied by a $1/\sqrt{2}$ for every
pair of identical particles in the initial or final state.

We can classify two-particle initial and final states 
according to their total electric charge $Q$ and 
hypercharge $Y$, following the approach in \cite{Bento:2017eti}.  
In addition to $Q$ and $Y$, any scattering process involving 
$SU(2)$ doublets must also conserve total isospin, $T$.
Following the results derived in Appendix \ref{app:REDUNDANT},
we list in Table~\ref{basistable} the minimal
set of independent two-particle states,
labeled by the quantum numbers $|Q, Y, T\rangle$.
Throughout we use the notation
\begin{eqnarray}
\phi^+_{[i}\, \phi^0_{j]} &\equiv&
\frac{1}{\sqrt{2}}\left(\phi^+_i\, \phi^0_j - \phi^+_j\, \phi^0_i\right)	,\\
\Phi_i\, \Phi_j^* &\equiv&
\frac{1}{\sqrt{2}}\left(\phi^+_i\, \phi^-_j + \phi^0_i\, \phi^{0*}_j\right).
\label{def_PhiGrande}
\end{eqnarray}
\begin{table}[]
\centering
\caption{Basis of two-particle states labeled by $|Q, Y, T\rangle$.
This table had been reduced by including only non-redundant sets of states;
see Appendix~\ref{app:REDUNDANT} for more details.}
\renewcommand{\arraystretch}{1.5}
\setlength{\tabcolsep}{4pt}
\centering
\begin{tabular}{c | c |  c | c}
$|Q, Y, T\rangle$ & State & Conditions & Dimensionality\\
\hline\hline
$|2, 2, 0\rangle$						
&	$\varphi^+_i \varphi^+_j$	
& $i\leq j$
& $n_c(n_c+1)/2$
\\\hline
$|2, \frac{3}{2}, \frac{1}{2}\rangle$	
&	$\phi^+_i \varphi^+_j$		
& ---
& $n_c n_D$
\\\hline
$|2, 1, 1\rangle$						
&	$\phi^+_i \phi^+_j$		
& $i\leq j$
& $n_D(n_D+1)/2$
\\\hline
$|1, 1, 0\rangle$						
& $\left\{\phi^+_{[i}\, \phi^0_{j]} , \ \varphi^+_i \chi_j \right\}$	
& $\{ i < j,\, \textrm{---}\,  \}$
& $n_D(n_D-1)/2 + n_c n_n$
\\\hline
$|1, \frac{1}{2}, \frac{1}{2}\rangle$	
&	
$\left\{\phi^+_i \chi_j , \ \phi^{0*}_i \varphi^+_j \right\}$
&	---
& $n_D(n_c + n_n)$
\\\hline
$|1, 0, 1\rangle$						
&	$\phi^+_i \phi^{0*}_j$	
&   ---
&  $n_D^2$
\\\hline
$|0, 0, 0\rangle$						
&	$\left\{ \Phi_i\Phi_j^* , \  \varphi_i^+ \varphi_j^-,\ \chi_i \chi_j \right\}$
& $\{\, \textrm{---}\,  ,\, \textrm{---}\,  ,\, i\leq j \}$	
& $n_n(n_n+1)/2 + n_D^2 + n_c^2$
\\\hline
\end{tabular}
\label{basistable}
\end{table}

Looking at Table~\ref{basistable}, we notice that the 
non-redundant sets of states could be classified exclusively in terms of
$Y$ and $T$, as is done for the 2HDM in \cite{Ginzburg:2005dt}. 
On the other hand, Ref.~\cite{Bento:2017eti} advocated for $Q$ and $Y$
when considering all states,
since, as can be seen in Table~\ref{tab:basisQY} of Appendix~\ref{app:REDUNDANT},
there are states with different $Q$ for the same $(Y,T)$.
Nevertheless, when considering scalars in representations of $SU(2)\times U(1)$
other than the ones we use here, the labeling of two-particle states by $Q$, $Y$, and $T$
is the most convenient as it leads to the minimal set of scattering matrices.
Therefore, we find it best to make the bridge and consider $|Q, Y, T\rangle$.

Using Eqs.~\eqref{eq:potential}--\eqref{rrr}, and \eqref{UNI}, 
the scattering matrix elements for the relevant processes read
\begin{eqnarray}
16\pi \, a_0\left[ \varphi^+_a \varphi^+_b \to \varphi^+_c \varphi^+_d
\right]
&=&
4 \,N_{ab} N_{cd}\, \alpha_{ca,db},
\\ 
16\pi \, a_0\left[ \phi^+_a \varphi^+_b \to \phi^+_c \varphi^+_d
\right]
&=&
\delta_{ca,db},
\\
16\pi \, a_0\left[ \phi^+_a \phi^+_b \to \phi^+_c \phi^+_d
\right]
&=&
2N_{ab} N_{cd}\,\left( \lambda_{ca,db}+\lambda_{da,cb} \right),
\\
16\pi \, a_0\left[ \phi^+_{[a}\, \phi^0_{b]}
\to
\phi^+_{[c}\, \phi^0_{d]}
\right]
&=&
2\left( \lambda_{ca,db}-\lambda_{da,cb} \right),
\\
16\pi \, a_0\left[ \phi^+_{[a}\, \phi^0_{b]}
\to
\varphi^+_c \chi_d
\right]
&=&
2\sqrt{2}\,i\,\kappa_{ba,cd},
\\
16\pi \, a_0\left[ \varphi^+_a \chi_b \to \varphi^+_c \chi_d
\right]
&=&
2\zeta_{ca,bd},
\\
16\pi \, a_0\left[ \phi^+_a \chi_b \to \phi^+_c \chi_d
\right]
&=&
2\gamma_{ca,bd},
\\
16\pi \, a_0\left[ \phi^+_a \chi_b \to \phi^{0*}_c \varphi^+_d
\right]
&=&
2i\kappa_{ca,db},
\\
16\pi \, a_0\left[ \phi^{0*}_a \varphi^+_b \to \phi^{0*}_c \varphi^+_d
\right]
&=&
\delta_{ac,db},
\\
16\pi \, a_0\left[ \phi^+_a \phi^{0*}_b \to \phi^+_c \phi^{0*}_d
\right]
&=&
2\lambda_{ca,bd},
\\
16\pi \, a_0\left[ \Phi_a\, \Phi_b^*
\to 
\Phi_c\, \Phi_d^*
\right]
&=&
4\lambda_{ba,cd} + 2\lambda_{ca,bd},
\\
16\pi \, a_0\left[ \Phi_a\, \Phi_b^*
\to 
\varphi^+_c \varphi^-_d
\right]
&=&
\sqrt{2}\delta_{ba,cd},
\\
16\pi \, a_0\left[ \Phi_a\, \Phi_b^*
\to 
\chi_{c}\, \chi_{d}
\right]
&=&
2\sqrt{2}N_{cd}\,\gamma_{ba,cd},
\\
16\pi \, a_0\left[ \varphi^+_a \varphi^-_b
\to 
\varphi^+_c \varphi^-_d
\right]
&=&
4\alpha_{ba,cd},
\\
16\pi \, a_0\left[ \varphi^+_a \varphi^-_b
\to 
\chi_{c}\, \chi_{d}
\right]
&=&
2\,N_{cd}\,\zeta_{ba,cd},
\\
16\pi \, a_0\left[ \chi_{a}\, \chi_{b}
\to 
\chi_{c}\, \chi_{d}
\right]
&=&
24\,N_{ab} N_{cd}\,\beta_{ab,cd}.
\end{eqnarray}
Below, we study specific models,
showing the scattering matrices $M_{|Q,Y,T\rangle}$
and, when an analytical expression is possible,
the corresponding eigenvalues. 
Note that in Ref.~\cite{Bento:2022vsb}
a method is proposed based on principal minors,
that forgoes diagonalization.
In specific numerical simulations,
this is likely to be computationally preferable
in all models dealing with large scattering matrices.

\section{\label{sec:Math}Mathematica notebook}

Deriving partial-wave unitarity bounds for 
different models is a rather repetitive task. 
As stated, for a model specified by definite values of 
$n_D$, $n_c$, and $n_n$, 
the procedure requires assembling a scattering matrix for each state 
$|Q, Y, T\rangle$ listed in Table~\ref{basistable}, 
computing the corresponding eigenvalues, 
and verifying that they remain below $8\pi$. 
In models with additional flavour symmetries, 
all quartic terms forbidden by those symmetries 
must also be set to zero.

To optimize this process and reduce the burden on the 
high-energy physics phenomenology community, 
we have developed the \texttt{Mathematica} 
tool {\tt BounDS} that automates the calculation. 
The user simply has to specify the values of 
$n_D$, $n_c$, and $n_n$, and, 
if needed, the transformation properties of the fields 
under additional flavour and/or generalized CP symmetries which may either be discrete or continuous. 
{\tt BounDS} carries out the necessary steps to:
\begin{itemize}
	\item Compute the set of all linearly independent quartic couplings allowed by  the symmetries.
	\item Calculate the 7 independent scattering matrices $M_{|Q,Y,T\rangle}$.
	\item Block-diagonalize the scattering matrices by swapping rows and columns.
	\item Output the quartic part of the scalar potential and scattering matrices in \LaTeX \ form.
	\item Output closed expressions for the eigenvalues of the scattering matrices, when possible.
\end{itemize}
Additional details can be found in Appendix~\ref{app:NOTEBOOK}
and the notebook {\tt BounDS} can be downloaded from:
\begin{center}
\linkGIT
\end{center}

The results of the next section are derived from this notebook.

\section{Perturbative Unitarity Bounds for particular cases}
\label{particularcases}

We now proceed to study perturbative unitarity bounds in specific
scenarios by specifying the values
of $n_D$, $n_c$, and $n_n$,
as well as any additional symmetries.
The cases studied explicitly in this work are summarized
in Table~\ref{models}.
\begin{table}[h]
\centering
\caption{List of models presented as examples in this article.}
\renewcommand{\arraystretch}{1.3}
\setlength{\tabcolsep}{4pt}
\centering
\begin{tabular}{c | c | c | c | c}
$n_D$ & $n_n$ & $n_c$ & Symmetries & Ref.\\
\hline\hline
1 & 0 & 0 & --- &\cite{Lee:1977eg}
\\
\hline
1 & 1 & 0 & --- &
\\
\hline
2 & 0 & 0 & $\mathbb{Z}_2$ & \cite{Ginzburg:2005dt}
\\
\hline
2 & 0 & 0 & --- & \cite{Ginzburg:2005dt}
\\
\hline
1 & 2 & 0 & --- & \cite{Muhlleitner:2020wwk} generalized
\\
\hline
2 & 1 & 0 & $\mathbb{Z}_2$ & 
\\
\hline
2 & 2 & 0 & $\mathbb{Z}_2 \otimes \mathbb{Z}'_2$ & \cite{Boto:2025mmn}
\\
\hline
2 & 1 & 1 & --- & 
\\
\hline
3 & 0 & 0 &  $\mathbb{Z}_3$ & \cite{Bento:2017eti}
\\
\hline
\end{tabular}
\label{models}
\end{table}

\subsection{The Standard Model}

In the Standard Model, the scalar sector consists of a single $SU(2)$ doublet
$\Phi_1$.
Consequently, the quartic part of the scalar potential simply reads
\be
V_4 = \lambda_{11,11} \left(\Phi_1^\dagger \Phi_1\right)^2.
\label{eq:potentialsm}
\ee

\subsubsection{Scattering Matrices}

For this minimal scalar content, there are only
three non-zero scattering matrices.
These matrices have rank-1 and take the following form:
\be
M_{|1, 0, 1\rangle} = M_{|2, 1, 1\rangle} = 
2\lambda_{11,11},
\label{SM111}
\ee
\be
M_{|0, 0, 0\rangle} = 
6\lambda_{11,11}.
\label{SM222}
\ee

\subsubsection{Perturbative Unitarity Bounds}

Applying the partial-wave unitarity condition
of Eq.~\eqref{UNI} to the eigenvalues
of these scattering matrices, immediately leads to
\be
|2\lambda_{11,11}| \leq 8\pi, \quad |6\lambda_{11,11}|
\leq 8\pi \quad \Longrightarrow \quad \lambda_{11,11} \leq \frac{4\pi}{3}.
\ee

In order to facilitate the comparison of our results with those of
Ref.~\cite{Bento:2017eti}, it is useful to provide a short
\textit{dictionary} of notations. Table~\ref{notationSM} lists the
correspondence between the coupling employed in this work and the one
used in Ref.~\cite{Bento:2017eti}.
\bt
\centering
\caption{Comparison of coupling notation.}
\renewcommand{\arraystretch}{1.3}
\setlength{\tabcolsep}{4pt}
\centering
\begin{tabular}{lccc}
\hline
\hline
 & \textbf{Term} & \textbf{Our Notation} & \textbf{Notation in \cite{Bento:2017eti}} \\
\hline
& $(\Phi_1^\dagger\Phi_1)^2$ & $\lambda_{11,11}$ & $\lambda$ \\
\hline
\end{tabular}
\label{notationSM}
\et
This recovers the classic results of \cite{Lee:1977eg}.

\subsection{1 Scalar Doublet and 1 Neutral Scalar Singlet}

We now consider an extension of the SM where,
besides the $SU(2)$ scalar doublet $\Phi_1$, we add
one neutral scalar singlet $\chi_1$.
In this case, the quartic part of the scalar potential takes the form
\be
V_4 = \lambda_{11,11} \left(\Phi_1^\dagger \Phi_1\right)^2
+ \beta_{11,11}\, \chi_1^4
+ \gamma_{11,11} \left(\Phi_1^\dagger \Phi_1\right)\chi_1^2.
\label{eq:potential_1d1s}
\ee

\subsubsection{Scattering Matrices}

The set of non-zero scattering matrices 
for this model are given by:
\be
M_{|1, 0, 1\rangle} = M_{|2, 1, 1\rangle} =
2\lambda_{11,11},
\ee
\be
M_{|1, \frac{1}{2},\frac{1}{2}\rangle} =
2\gamma_{11,11},
\ee
\be
M_{|0,0,0\rangle} =
\bb
6 \lambda_{11,11} & 2 \gamma_{11,11} \\
 2 \gamma_{11,11} & 12 \beta_{11,11} \\
\eb.
\ee

\subsubsection{Perturbative Unitarity Bounds}

We apply the partial-wave unitarity condition
of Eq.~\eqref{UNI} to each of the eigenvalues 
of the zero partial-wave amplitude matrix
and find
\begin{eqnarray}
\left|
6\beta_{11,11} + 3\lambda_{11,11} \pm
\sqrt{9 \left(\lambda _{11,11}-2 \beta _{11,11}\right){}^2+4 \gamma _{11,11}^2}
\right|
& \leq & 8\pi,
\\
|2\lambda_{11,11}|
& \leq & 8\pi,
\\
|2\gamma_{11,11}|
& \leq & 8\pi.
\end{eqnarray}

\subsection{The $\mathbb{Z}_2$-Symmetric 2HDM}

We now consider a 2HDM model with a discrete $\mathbb{Z}_2$
symmetry that acts on the two scalar $SU(2)$ doublets,
$\Phi_1$ and $\Phi_2$, as
\be
\Phi_1 \to \Phi_1, \qquad \Phi_2 \to -\Phi_2.
\ee
This case illustrates the use of our {\tt Mathematica} 
program in the presence of symmetries.
Under this symmetry, the quartic part of the scalar potential reads
\begin{eqnarray}
V_4^{\mathbb{Z}_2} &=&
 \lambda_{11,11} \left(\Phi_1^\dagger \Phi_1\right)^2
+  \lambda_{22,22} \left(\Phi_2^\dagger \Phi_2\right)^2
+ 2\lambda_{11,22} \left(\Phi_1^\dagger \Phi_1\right)\left(\Phi_2^\dagger \Phi_2\right)
+ 2\lambda_{12,21} \left(\Phi_1^\dagger \Phi_2\right)\left(\Phi_2^\dagger \Phi_1\right)
\\
&&  
\ +\  \lambda_{12,12}\left(\Phi_1^\dagger \Phi_2\right)^2
+ \lambda_{12,12}^*\left(\Phi_2^\dagger \Phi_1\right)^2 
\\[0.5em]
 &=&
 \frac{\lambda_1}{2} \left(\Phi_1^\dagger \Phi_1\right)^2
+ \frac{\lambda_2}{2} \left(\Phi_2^\dagger \Phi_2\right)^2
+ \lambda_3 \left(\Phi_1^\dagger \Phi_1\right)\left(\Phi_2^\dagger \Phi_2\right)
+ \lambda_4 \left(\Phi_1^\dagger \Phi_2\right)\left(\Phi_2^\dagger \Phi_1\right)
+ \frac{\lambda_5}{2} \left[ \left(\Phi_1^\dagger \Phi_2\right)^2
+ \text{h.c.} \right],
\label{2hdmz2new}
\end{eqnarray}
where, in Eq.~\eqref{2hdmz2new} we have employed the
standard notation of Ref.~\cite{Ginzburg:2005dt} and,
without loss of generality, choose \(\lambda_5\) to be real.
The mapping between the two notations is given in 
Table~\ref{tab:2HDMnotation}.
\bt
\centering
\caption{Comparison of coupling notation.}
\renewcommand{\arraystretch}{1.3}
\setlength{\tabcolsep}{4pt}
\centering
\begin{tabular}{lccc}
\hline
 & \textbf{Term} & \textbf{Our Notation} & \textbf{Notation in \cite{Ginzburg:2005dt}} \\
\hline
\hline
& $(\Phi_1^\dagger \Phi_1)^2$ & $\lambda_{11,11}$ & $\frac{\lambda_1}{2}$ \\
\hline
& $(\Phi_2^\dagger \Phi_2)^2$ & $\lambda_{22,22}$ & $\frac{\lambda_2}{2}$ \\
\hline
& $(\Phi_1^\dagger \Phi_1)(\Phi_2^\dagger \Phi_2)$ & $2\lambda_{11,22}$ & $\lambda_3$ \\
\hline
& $(\Phi_1^\dagger \Phi_2)(\Phi_2^\dagger \Phi_1)$ & $2\lambda_{12,21}$ & $\lambda_4$ \\
\hline
& $(\Phi_1^\dagger \Phi_2)^2$ & $\lambda_{12,12}$ & $\frac{\lambda_5}{2}$ \\
\hline
& $(\Phi_2^\dagger \Phi_1)^2$ & $\lambda_{12,12}^*$ & $\frac{\lambda_5}{2}$ \\
\hline
\end{tabular}
\label{tab:2HDMnotation}
\et

\subsubsection{Scattering Matrices}
\label{z22hdm}

The set of non-zero scattering matrices 
for this model reads
\be
M_{|2, 1, 1\rangle} = 
\bb
\lambda_{1} & \lambda_{5} & 0 \\
\lambda_{5} & \lambda_{2} & 0 \\
0 & 0 & \lambda_{3} + \lambda_{4}
\eb,
\ee
\be
M_{|1,0,1\rangle} =
\bb
\lambda_{1} & \lambda_{4} & 0 & 0 \\
\lambda_{4} & \lambda_{2} & 0 & 0 \\
0 & 0 & \lambda_{3} & \lambda_{5} \\
0 & 0 & \lambda_{5} & \lambda_{3}
\eb,
\ee
\be
M_{|0,0,0\rangle} =
\bb
3\lambda_{1} & 2\lambda_{3} + \lambda_{4} & 0 & 0 \\
2\lambda_{3} + \lambda_{4} & 3\lambda_{2} & 0 & 0 \\
0 & 0 & \lambda_{3} + 2\lambda_{4} & 3\lambda_{5} \\
0 & 0 & 3\lambda_{5} & \lambda_{3} + 2\lambda_{4}
\eb,
\ee
\be
M_{|1,1,0\rangle} =
\lambda_3-\lambda_4.
\ee

\subsubsection{Eigenvalues}

By imposing the unitarity condition
of Eq.~\eqref{UNI} to each of the eigenvalues 
of the zero partial-wave amplitude matrix
we derive
\ba
\left|  \lambda_3 \pm \lambda_4 \right| &\leq & 8\pi,
\label{Z2-2HDM__1}
\\
\left|  \lambda_3 \pm \lambda_5 \right| &\leq & 8\pi,
\\
\left|  \lambda_3 + 2 \lambda_4 \pm 3 \lambda_5\right| &\leq & 8\pi,
\\
\frac{1}{2}
\left|\lambda_1+\lambda_2
\pm \sqrt{\left(\lambda_1-\lambda_2\right)^{2}
+ 4 \lambda_4^2 }\right| &\leq & 8\pi,
\\
\frac{1}{2}
\left| \lambda_1+\lambda_2
\pm \sqrt{\left(\lambda_1-\lambda_2\right)^{2}
+4 \lambda_5^2} \right| &\leq & 8\pi,
\\
\frac{1}{2}
\left| 3 \lambda_1+3 \lambda_2
\pm \sqrt{9 \left(\lambda_1-\lambda_2\right)^{2}
+ 4 \left(2 \lambda_3+\lambda_4\right)^{2}} \right|
&\leq & 8\pi.
\label{Z2-2HDM}
\ea
We find that our results are in agreement 
with those derived in \cite{Ginzburg:2005dt}.

\subsection{The general 2HDM}

In this case, we consider the most general scalar potential involving
two $SU(2)$ doublets, \(\Phi_1\) and \(\Phi_2\).
The quartic part of the potential is given by:
\begin{eqnarray}
V_4 &=& 
\lambda_{11,11} \left(\Phi_1^\dagger \Phi_1\right)^2
+ \lambda_{22,22} \left(\Phi_2^\dagger \Phi_2\right)^2
+ 2\lambda_{11,22} \left(\Phi_1^\dagger \Phi_1\right)\left(\Phi_2^\dagger \Phi_2\right)
+ 2\lambda_{12,21} \left(\Phi_1^\dagger \Phi_2\right)\left(\Phi_2^\dagger \Phi_1\right) 
+ \nonumber \\
&&\left[ \lambda_{12,12} \left(\Phi_1^\dagger \Phi_2\right)^2 
+ 2\lambda_{11,12} \left(\Phi_1^\dagger \Phi_1\right)\left(\Phi_1^\dagger \Phi_2\right)
+ 2\lambda_{12,22} \left(\Phi_1^\dagger \Phi_2\right)\left(\Phi_2^\dagger \Phi_2\right)
+ \text{h.c.} \right]\\[0.5em]
&=&\frac{\lambda_1}{2} \left(\Phi_1^\dagger \Phi_1\right)^2
+ \frac{\lambda_2}{2} \left(\Phi_2^\dagger \Phi_2\right)^2
+ \lambda_3 \left(\Phi_1^\dagger \Phi_1\right)\left(\Phi_2^\dagger \Phi_2\right)
+ \lambda_4 \left(\Phi_1^\dagger \Phi_2\right)\left(\Phi_2^\dagger \Phi_1\right) 
+ \nonumber \\
&&\left[ \frac{\lambda_5}{2} \left(\Phi_1^\dagger \Phi_2\right)^2 
+ \lambda_6 \left(\Phi_1^\dagger \Phi_1\right)\left(\Phi_1^\dagger \Phi_2\right)
+ \lambda_7 \left(\Phi_1^\dagger \Phi_2\right)\left(\Phi_2^\dagger \Phi_2\right)
+ \text{h.c.} \right],
\label{dewcerwcerc}
\end{eqnarray}
where, in Eq.~\eqref{dewcerwcerc} we have employed, once again, the
standard notation of Ref.~\cite{Ginzburg:2005dt}.
The mapping between the two notations is given in 
Table~\ref{tab:2HDMnotation_general}.
\bt
\centering
\caption{Comparison of coupling notation.}
\renewcommand{\arraystretch}{1.3}
\setlength{\tabcolsep}{4pt}
\centering
\begin{tabular}{lccc}
\hline
 & \textbf{Term} & \textbf{Our Notation} & \textbf{Notation in \cite{Ginzburg:2005dt}} \\
\hline
\hline
& $(\Phi_1^\dagger \Phi_1)^2$ & $\lambda_{11,11}$ & $\frac{\lambda_1}{2}$ \\
\hline
& $(\Phi_2^\dagger \Phi_2)^2$ & $\lambda_{22,22}$ & $\frac{\lambda_2}{2}$ \\
\hline
& $(\Phi_1^\dagger \Phi_1)(\Phi_2^\dagger \Phi_2)$ & $2\lambda_{11,22}$ & $\lambda_3$ \\
\hline
& $(\Phi_1^\dagger \Phi_2)(\Phi_2^\dagger \Phi_1)$ & $2\lambda_{12,21}$ & $\lambda_4$ \\
\hline
& $(\Phi_1^\dagger \Phi_2)^2$ & $\lambda_{12,12}$ & $\frac{\lambda_5}{2}$ \\
\hline
& $(\Phi_1^\dagger \Phi_1)(\Phi_1^\dagger \Phi_2)$ & 
$2\lambda_{11,12}$ & $\lambda_6$ \\
\hline
& $(\Phi_1^\dagger \Phi_2)(\Phi_2^\dagger \Phi_2)$ & 
$2\lambda_{12,22}$ & $\lambda_7$ \\
\hline
\end{tabular}
\label{tab:2HDMnotation_general}
\et

\subsubsection{Scattering Matrices}

The set of non-zero scattering matrices 
for this model reads
\be
M_{|2,1,1\rangle} =
\bb
\lambda_1 & \sqrt{2}\,\lambda_6 & \lambda_5 \\
\sqrt{2}\,\lambda_6^* & \lambda_3 + \lambda_4 & \sqrt{2}\,\lambda_7 \\
\lambda_5^* & \sqrt{2}\,\lambda_7^* & \lambda_2
\eb,
\ee
\be
M_{|1,0,1\rangle} =
\bb
\lambda_1 & \lambda_6^* & \lambda_6 & \lambda_4 \\
\lambda_6 & \lambda_3 & \lambda_5 & \lambda_7 \\
\lambda_6^* & \lambda_5^* & \lambda_3 & \lambda_7^* \\
\lambda_4 & \lambda_7^* & \lambda_7 & \lambda_2
\eb,
\ee
\be
M_{|0,0,0\rangle} =
\bb
3 \lambda_1 & 3 \lambda_6^* & 3 \lambda_6 & 2 \lambda_3 + \lambda_4 \\
3 \lambda_6 & \lambda_3 + 2 \lambda_4 & 3 \lambda_5 & 3 \lambda_7 \\
3 \lambda_6^* & 3 \lambda_5^* & \lambda_3 + 2 \lambda_4 & 3 \lambda_7^* \\
2 \lambda_3 + \lambda_4 & 3 \lambda_7^* & 3 \lambda_7 & 3 \lambda_2
\eb,
\ee
\be
M_{|1,1,0\rangle} =
\lambda_3-\lambda_4.
\ee
Our scattering matrices agree with those in \cite{Ginzburg:2005dt}.
If we set $\lambda_6 = \lambda_7 = 0$ (along with their complex conjugates),
we re-obtain, as expected,
the results presented in Section~\ref{z22hdm}.

\subsubsection{Perturbative Unitarity Bounds}

The eigenvalues of the scattering matrices in this case are, in general,
too complex to write down in closed form.
However, some can be computed analytically,
and perturbative unitarity bounds are imposed accordingly,
\be
|\lambda_3 - \lambda_4| \leq 8\pi.
\ee

\subsection{1 Scalar Doublet and 2 Neutral Scalar Singlets}

In this case, the most general quartic part of the scalar potential reads:
\begin{align}
V_4 =\ 
& \lambda_{11,11} \left( \Phi_1^\dagger \Phi_1 \right)^2
+ \beta_{11,11}\, \chi_1^4 + \beta_{22,22}\, \chi_2^4 
+ 6\, \beta_{11,22}\, \chi_1^2 \chi_2^2 + 4\, \beta_{11,12}\, \chi_1^3 \chi_2
+ 4\, \beta_{12,22}\, \chi_1\chi_2^3 \nonumber \\
& + \gamma_{11,11} \left( \Phi_1^\dagger \Phi_1 \right) \chi_1^2 
+ \gamma_{11,22} \left( \Phi_1^\dagger \Phi_1 \right) \chi_2^2 
+ 2\, \gamma_{11,12} \left( \Phi_1^\dagger \Phi_1 \right) \chi_1 \chi_2,
\label{poooe3}
\end{align}
where $\Phi_1$ denotes the scalar doublet and $\chi_1$ and $\chi_2$ 
are the real scalar singlets.
We have used the relations in Eqs.~\eqref{propsbeta} and \eqref{propsgamma}
to simplify the quartic part of the potential.

\subsubsection{Scattering Matrices}
The set of non-zero scattering matrices 
for this model reads
\be
M_{|1,0,1\rangle}= M_{|2,1,1\rangle} = 
2 \lambda_{11,11},
\ee
\be
M_{|1,\frac{1}{2}, \frac{1}{2}\rangle} =
\bb
2 \gamma_{11,11} & 2 \gamma_{11,12} \\
2 \gamma_{11,12} & 2 \gamma_{11,22}
\eb,
\ee
\be
M_{|0,0,0\rangle} =
\left[
\begin{array}{cccc}
 6 \lambda _{11,11} & 2 \gamma
   _{11,11} & 2 \sqrt{2} \gamma
   _{11,12} & 2 \gamma _{11,22} \\
 2 \gamma _{11,11} & 12 \beta
   _{11,11} & 12 \sqrt{2} \beta
   _{11,12} & 12 \beta _{11,22} \\
 2 \sqrt{2} \gamma _{11,12} & 12
   \sqrt{2} \beta _{11,12} & 24
   \beta _{11,22} & 12 \sqrt{2}
   \beta _{12,22} \\
 2 \gamma _{11,22} & 12 \beta
   _{11,22} & 12 \sqrt{2} \beta
   _{12,22} & 12 \beta _{22,22} \\
\end{array}
\right].
\ee

\subsubsection{Perturbative Unitarity Bounds}

Again, certain eigenvalues of the scattering matrices are
too complicated to be expressed in closed form.
However, others can be written analytically,
and, for those, the corresponding perturbative unitarity bounds are then applied, as:
\begin{align}
\left| 
\gamma _{11,11}+\gamma _{11,22} \pm \sqrt{4 \gamma _{11,12}^2+\left(\gamma _{11,11}-\gamma _{11,22}\right){}^2}
 \right|
&\leq 8\pi,\\
|2 \lambda_{11,11}| &\leq 8\pi, \\
\end{align}

In order to compare our results with previous work,
we consider the model studied in Ref.~\cite{Muhlleitner:2020wwk},
where the SM is extended by a neutral complex scalar singlet.
In their notation, the
quartic part of the scalar potential reads
\ba
V_4 = \frac{\lambda}{4} \left(\Phi_1^\dagger\Phi_1\right)^2
+ \frac{\delta_2}{2}\left(\Phi_1^\dagger\Phi_1\right)|S|^2
+ \frac{d_2}{2}|S|^4, 
\label{asadce}
\ea
where
$S = \frac{\chi_1 + i \chi_2}{\sqrt{2}}$.
Notice that the case with two real singlets in Eq.~\eqref{poooe3} is more general than
the case with one complex singlet in Eq.~\eqref{asadce}.
But, since the complex scalar singlet can be decomposed into
two real scalar singlet fields, $\chi_1$ and $\chi_2$,
we find that our results reproduce those of \cite{Muhlleitner:2020wwk},
when we set $\gamma_{11,12} = \beta_{11,12} = \beta_{12,22} = 0$,
\textit{i.e.} when we impose an additional $\mathbb{Z}_2$
that acts as $\chi_2\to -\chi_2$.\footnote{Equivalently,
one could choose $S = \frac{\chi_2 + i \chi_1}{\sqrt{2}}$ and impose
the $\mathbb{Z}_2$ symmetry on $\chi_1$.
This procedure is generalizable to models with any number
of complex neutral singlet scalars.}
By matching the terms in Eq.~\eqref{asadce}
to those in Eq.~\eqref{poooe3}, we derive 
the mapping between the two notations that is
provided in Table~\ref{notation1d2s}.
\bt
\centering
\caption{Comparison of coupling notation.}
\renewcommand{\arraystretch}{1.3}
\setlength{\tabcolsep}{4pt}
\centering
\begin{tabular}{lccc}
\hline
 & \textbf{Term} & \textbf{Our Notation} & \textbf{Notation in \cite{Muhlleitner:2020wwk}} \\
\hline\hline
& $(\Phi_1^\dagger\Phi_1)^2$ & $\lambda_{11,11}$ & $\frac{\lambda}{4}$ \\
\hline
& $\chi_1^4$ & $\beta_{11,11}$ & $\frac{d_2}{16}$ \\
\hline
& $\chi_2^4$ & $\beta_{22,22}$ & $\frac{d_2}{16}$ \\
\hline
& $\chi_1^2\chi_2^2$ & $6\beta_{11,22}$ & $\frac{d_2}{8}$ \\
\hline
& $\chi_1^3\chi_2$ & $4\beta_{11,12}$ & 0 \\
\hline
& $\chi_1\chi_2^3$ & $4\beta_{12,22}$ & 0 \\
\hline
& $(\Phi_1^\dagger\Phi_1)\chi_1^2$ & $\gamma_{11,11}$ & $\frac{\delta_2}{4}$ \\
\hline
& $(\Phi_1^\dagger\Phi_1)\chi_2^2$ & $\gamma_{11,22}$ & $\frac{\delta_2}{4}$ \\
\hline
& $(\Phi_1^\dagger\Phi_1)\chi_1\chi_2$ & $2\gamma_{11,12}$ & $0$ \\
\hline
\end{tabular}
\label{notation1d2s}
\et

\subsection{2 Scalar Doublets and 1 Neutral Scalar Singlet with a $\mathbb{Z}_2$ symmetry}

We now consider a 2HDM with one additional
neutral scalar singlet $\chi_1$.
We impose a $\mathbb{Z}_2$ symmetry
that acts on the fields as
\be
\Phi_1 \to \Phi_1, \qquad \Phi_2 \to -\Phi_2,
\qquad \chi_1 \to \chi_1.
\ee
Under this symmetry, and using Eqs.~\eqref{propslambda}, \eqref{propsbeta}, and
\eqref{propsgamma}, the scalar potential simplifies to:
\begin{align}
V_4 =\ 
& \lambda_{11,11} \left(\Phi_1^\dagger \Phi_1\right)^2 +
\lambda_{22,22} \left(\Phi_2^\dagger \Phi_2\right)^2 +
2\lambda_{11,22} \left(\Phi_1^\dagger \Phi_1\right)\left(\Phi_2^\dagger \Phi_2\right) +
2\lambda_{12,21} \left(\Phi_1^\dagger \Phi_2\right)\left(\Phi_2^\dagger \Phi_1\right) \nonumber \\
& +
\left[
\lambda_{12,12} \left(\Phi_1^\dagger \Phi_2\right)^2 + \text{h.c.}
\right] +
\beta_{11,11} \chi_1^4 +
\gamma_{11,11} \left(\Phi_1^\dagger \Phi_1\right) \chi_1^2 +
\gamma_{22,11} \left(\Phi_2^\dagger \Phi_2\right) \chi_1^2.
\end{align}

\subsubsection{Scattering Matrices}
The set of non-zero scattering matrices 
for this model reads
\be
M_{|1,\frac{1}{2},\frac{1}{2}\rangle} =
\bb
2 \gamma_{11,11} & 0 \\
0 & 2 \gamma_{22,11}
\eb,
\ee
\be
M_{|2,1,1\rangle} =
\bb
2\lambda_{11,11} & 2\lambda_{12,12} & 0 \\
2\lambda^*_{12,12} & 2\lambda_{22,22} & 0 \\
0 & 0 & 2(\lambda_{11,22} + \lambda_{12,21})
\eb,
\ee
\be
M_{|1,0,1\rangle} =
\bb
2\lambda_{11,11} & 2\lambda_{12,21} & 0 & 0 \\
2\lambda_{12,21} & 2\lambda_{22,22} & 0 & 0 \\
0 & 0 & 2\lambda_{11,22} & 2\lambda_{12,12} \\
0 & 0 & 2\lambda^*_{12,12} & 2\lambda_{11,22}
\eb,
\ee
\be
M_{|0,0,0\rangle} =
\left[
\begin{array}{ccccc}
 6 \lambda _{11,11} & 2 \left(2
   \lambda _{11,22}+\lambda
   _{12,21}\right) & 2 \gamma
   _{11,11} & 0 & 0 \\
 2 \left(2 \lambda _{11,22}+\lambda
   _{12,21}\right) & 6 \lambda
   _{22,22} & 2 \gamma _{22,11} & 0
   & 0 \\
 2 \gamma _{11,11} & 2 \gamma
   _{22,11} & 12 \beta _{11,11} & 0
   & 0 \\
 0 & 0 & 0 & 2 \left(\lambda
   _{11,22}+2 \lambda
   _{12,21}\right) & 6 \lambda
   _{12,12} \\
 0 & 0 & 0 & 6 \lambda _{12,12}^* &
   2 \left(\lambda _{11,22}+2
   \lambda _{12,21}\right) \\
\end{array}
\right],
\ee
\be
M_{|1,1,0\rangle} =
2 \left( \lambda_{11,22} - \lambda_{12,21} \right).
\ee

\subsubsection{Perturbative Unitarity Bounds}

Some eigenvalues are too complicated to be written in closed form.
Therefore, we show only the results of imposing partial-wave unitarity on the
remaining eigenvalues of the zero partial-wave amplitude matrix,
\begin{align}
|2 \gamma_{11,11}| &\leq 8\pi, \\[2mm]
|2 \gamma_{22,11}| &\leq 8\pi, \\[2mm]
2 | \lambda_{11,22} \pm \lambda_{12,21}| &\leq 8\pi, \\[2mm]
2 \left| \lambda_{11,22} \pm |\lambda_{12,12}| \right| &\leq 8\pi, \\[2mm]
\left| \lambda_{11,11} + \lambda_{22,22} \pm 
\sqrt{(\lambda_{11,11} - \lambda_{22,22})^2 + 4 \lambda_{12,21}^2} \right| &\leq 8\pi, \\[2mm]
\left| \lambda_{11,11} + \lambda_{22,22} \pm 
\sqrt{(\lambda_{11,11} - \lambda_{22,22})^2 + 4 |\lambda_{12,12}|^2} \right| &\leq 8\pi,\\[2mm]
2 \left|\lambda
   _{11,22}+2 \lambda
   _{12,21} \pm3 |\lambda_{12,12}|^2\right| &\leq 8\pi.
\end{align}

\subsection{2 Scalar Doublets with a $\mathbb{Z}_2$ symmetry
+ 2 Neutral Scalar Singlets with a $\mathbb{Z'}_2$ symmetry}

We now consider a model with two $SU(2)$ doublets and
two neutral scalar singlets.
We impose two independent $\mathbb{Z}_2$ symmetries
that act on these fields as
\begin{eqnarray}
\mathbb{Z}_2:&&
\Phi_1 \to \Phi_1\quad \Phi_2 \to -\Phi_2,
\quad \chi_1 \to \chi_1,\quad \chi_2 \to \chi_2,\\
\mathbb{Z}_2^\prime:&&
\Phi_1 \to \Phi_1\quad \Phi_2 \to \Phi_2,
\quad \chi_1 \to -\chi_1,\quad \chi_2 \to -\chi_2,
\end{eqnarray}

Under these symmetries, the quartic 
part of the scalar potential becomes:
\begin{align}
V_4 &= \lambda_{11,11} \left(\Phi_1^\dagger \Phi_1\right)^2 +
\lambda_{22,22} \left(\Phi_2^\dagger \Phi_2\right)^2 +
2\lambda_{11,22} \left(\Phi_1^\dagger \Phi_1\right)\left(\Phi_2^\dagger \Phi_2\right) +
2\lambda_{12,21} \left(\Phi_1^\dagger \Phi_2\right)\left(\Phi_2^\dagger \Phi_1\right) \nonumber \\
&\quad + \left[ \lambda_{12,12} \left(\Phi_1^\dagger \Phi_2 \right)^2 + \text{h.c.} \right]
+ \beta_{11,11}\, \chi_1^4 + \beta_{22,22}\, \chi_2^4
+ 6\beta_{11,22}\, \chi_1^2 \chi_2^2
+ 4\beta_{11,12}\, \chi_1^3 \chi_2
+ 4\beta_{12,22}\, \chi_2^3 \chi_1 \nonumber \\
&\quad + \gamma_{11,11} \left(\Phi_1^\dagger \Phi_1\right) \chi_1^2
+ \gamma_{22,11} \left(\Phi_2^\dagger \Phi_2\right) \chi_1^2
+ \gamma_{11,22} \left(\Phi_1^\dagger \Phi_1\right) \chi_2^2
+ 2\gamma_{11,12} \left(\Phi_1^\dagger \Phi_1\right) \chi_1 \chi_2 \nonumber \\
&\quad + 2\gamma_{22,12} \left(\Phi_2^\dagger \Phi_2\right) \chi_1 \chi_2
+ \gamma_{22,22} \left(\Phi_2^\dagger \Phi_2\right) \chi_2^2,
\end{align}
where we have simplified the quartic couplings 
using the relations in Eqs.~\eqref{propslambda} and \eqref{propsbeta}.

\subsubsection{Scattering Matrices}
The set of non-zero scattering matrices 
for this model reads
\be
M_{|1, \frac{1}{2}, \frac{1}{2}\rangle} =
\left[
\begin{array}{cccc}
 2 \gamma _{22,11} & 2 \gamma
   _{22,12} & 0 & 0 \\
 2 \gamma _{22,12} & 2 \gamma
   _{22,22} & 0 & 0 \\
 0 & 0 & 2 \gamma _{11,22} & 2
   \gamma _{11,12} \\
 0 & 0 & 2 \gamma _{11,12} & 2
   \gamma _{11,11} \\
\end{array}
\right],
\ee
\be
M_{|2,1,1\rangle} = 
\bb
2\lambda_{11,11} & 2\lambda_{12,12} & 0 \\
2\lambda_{12,12}^* & 2\lambda_{22,22} & 0 \\
0 & 0 & 2\lambda_{11,22} + 2\lambda_{12,21}
\eb,
\ee
\be
M_{|1,0,1\rangle} =
\bb
2\lambda_{11,11} & 2\lambda_{12,21} & 0 & 0 \\
2\lambda_{12,21} & 2\lambda_{22,22} & 0 & 0 \\
0 & 0 & 2\lambda_{11,22} & 2\lambda_{12,12} \\
0 & 0 & 2\lambda^*_{12,12} & 2\lambda_{11,22}
\eb,
\ee
\be
M_{|1,1,0\rangle} =
2 \left( \lambda_{11,22} -  \lambda_{12,21} \right),
\ee
\be
M_{|0,0,0\rangle}=
\textrm{blkdiag}(A,B)
\label{1st_block}
\ee
where, here and henceforth, $\textrm{blkdiag}(A,B,\dots)$ refers to a block diagonal
matrix, whose entries are the matrices $A,B,\dots$.
The matrices $A$ and $B$ in Eq.~\eqref{1st_block} are given, respectively, by
\be
A=
\bb
24 \, \beta_{11,22} & 2 \sqrt{2} \, \gamma_{11,12}
& 2 \sqrt{2} \, \gamma_{22,12} & 12 \sqrt{2} \, \beta_{11,12}
& 12 \sqrt{2} \, \beta_{12,22}
\\
2 \sqrt{2} \, \gamma_{11,12} & 6 \, \lambda_{11,11}
& 2(2 \, \lambda_{11,22} +  \, \lambda_{12,21})
& 2 \, \gamma_{11,11} & 2\, \gamma_{11,22}
\\
2 \sqrt{2} \, \gamma_{22,12} & 2(2 \, \lambda_{11,22} +  \, \lambda_{12,21})
& 6 \, \lambda_{22,22} & 2 \, \gamma_{22,11}
& 2 \, \gamma_{22,22}
\\
12 \sqrt{2} \, \beta_{11,12} & 2 \, \gamma_{11,11} & 2 \, \gamma_{22,11}
& 12 \, \beta_{11,11} & 12 \, \beta_{11,22}
\\
12 \sqrt{2} \, \beta_{12,22} & 2 \, \gamma_{11,22} & 2 \, \gamma_{22,22}
& 12 \, \beta_{11,22} & 12 \, \beta_{22,22}
\eb,
\ee
\be
B=
\bb
2 (\, \lambda_{11,22} + 2 \, \lambda_{12,21}) & 6 \, \lambda_{12,12}
\\
6 \, \lambda_{12,12}^* & 2 (\, \lambda_{11,22} + 2 \, \lambda_{12,21})
\eb.
\ee

\subsubsection{Perturbative Unitarity Bounds}

As several eigenvalues are too involved to express analytically,
we show only the results of imposing partial-wave unitarity on the
remaining eigenvalues of the $s$-wave amplitude matrix,
that can be computed explicitly,
\begin{eqnarray}
2 \left|  \lambda_{11,22} \pm \lambda_{12,21} \right|
& \leq & 8\pi,
\\[2mm]
2 \left| \lambda_{11,22} \pm |\lambda_{12,12}| \right|
& \leq & 8\pi,
\\[2mm]
2 \left|\lambda_{11,22} + 2 \lambda_{12,21} \pm 3|\lambda_{12,12}| \right|
& \leq & 8\pi,
\\[2mm]
\left| \lambda_{11,11} + \lambda_{22,22}
\pm \sqrt{ (\lambda_{11,11} - \lambda_{22,22})^2 + 4 \lambda_{12,21}^2 }
\right|
&\leq & 8\pi,
\\[2mm]
\left| \lambda_{11,11} + \lambda_{22,22}
\pm \sqrt{ (\lambda_{11,11} - \lambda_{22,22})^2 + 4 |\lambda_{12,12}|^2 }
\right|
& \leq & 8\pi,
\\[2mm]
\left|
\gamma_{11,11} + \gamma_{11,22}
\pm \sqrt{4 \gamma
   _{11,12}^2+\left(\gamma
   _{11,11}-\gamma
   _{11,22}\right){}^2}
\right|
&\leq & 8\pi,
\\[2mm]
\left| \gamma_{22,11} + \gamma_{22,22}
\pm \sqrt{ 4 \gamma
   _{22,12}^2+\left(\gamma
   _{22,11}-\gamma
   _{22,22}\right){}^2} \right|
&\leq& 8\pi.
\end{eqnarray}
\bt
\centering
\caption{Comparison of coupling notation.}
\renewcommand{\arraystretch}{1.3}
\setlength{\tabcolsep}{4pt}
\centering
\begin{tabular}{lccc}
\hline
 & \textbf{Term} & \textbf{Our Notation} & \textbf{Notation of \cite{Boto:2025mmn}} \\
\hline\hline
& $(\Phi_1^\dagger \Phi_1)^2$ & $\lambda_{11,11}$ & $\tfrac{\lambda_1}{2}$ \\
\hline
& $(\Phi_2^\dagger \Phi_2)^2$ & $\lambda_{22,22}$ & $\tfrac{\lambda_2}{2}$ \\
\hline
& $(\Phi_1^\dagger \Phi_1)(\Phi_2^\dagger \Phi_2)$ & $2\lambda_{11,22}$ & $\lambda_3$ \\
\hline
& $(\Phi_1^\dagger \Phi_2)(\Phi_2^\dagger \Phi_1)$ & $2\lambda_{12,21}$ & $\lambda_4$ \\
\hline
& $(\Phi_1^\dagger \Phi_2)^2$ & $\lambda_{12,12}$ & $\lambda_5$ \\
\hline
& $(\Phi_2^\dagger \Phi_1)^2$ & $\lambda_{12,12}^*$ & $\lambda_5$ \\
\hline
& $\chi_1^4$ & $\beta_{11,11}$ & $\tfrac{\lambda_6}{8}$ \\
\hline
& $\chi_2^4$ & $\beta_{22,22}$ & $\tfrac{\lambda_9}{8}$ \\
\hline
& $\chi_1^2 \chi_2^2$ & $6\beta_{11,22}$ & $\tfrac{\lambda_{10}}{4}$ \\
\hline
& $\chi_1^3 \chi_2$ & $4\beta_{11,12}$ & $\tfrac{\lambda_{13}}{6}$ \\
\hline
& $\chi_1 \chi_2^3$ & $4\beta_{12,22}$ & $\tfrac{\lambda_{14}}{6}$ \\
\hline
& $(\Phi_1^\dagger \Phi_1)\chi_1^2$ & $\gamma_{11,11}$ & $\tfrac{\lambda_7}{2}$ \\
\hline
& $(\Phi_2^\dagger \Phi_2)\chi_1^2$ & $\gamma_{22,11}$ & $\tfrac{\lambda_8}{2}$ \\
\hline
& $(\Phi_1^\dagger \Phi_1)\chi_2^2$ & $\gamma_{11,22}$ & $\tfrac{\lambda_{11}}{2}$ \\
\hline
& $(\Phi_2^\dagger \Phi_2)\chi_2^2$ & $\gamma_{22,22}$ & $\tfrac{\lambda_{12}}{2}$ \\
\hline
& $(\Phi_1^\dagger \Phi_1)\chi_1\chi_2$ & $2\gamma_{11,12}$ & $\tfrac{\lambda_{15}}{2}$ \\
\hline
& $(\Phi_2^\dagger \Phi_2)\chi_1\chi_2$ & $2\gamma_{22,12}$ & $\tfrac{\lambda_{16}}{2}$ \\
\hline
\end{tabular}
\label{notation2d2sZ2Z2}
\et
We verified our results against those reported in
\cite{Boto:2025mmn} and found them to be in agreement.
Table \ref{notation2d2sZ2Z2} summarizes the correspondence between
the couplings used in our work and those of \cite{Boto:2025mmn}.

\subsection{\label{sec:2D1N1C}2 Scalar Doublets + 1 Neutral Scalar Singlet + 1 Charged Scalar Singlet}

For a model with two $SU(2)$ scalar doublets $\Phi_1$ and $\Phi_2$,
one neutral scalar singlet $\chi_1$ and one
charged scalar singlet $\varphi^+_1$,
the quartic part of the scalar potential can be written as
\begin{align}
V_4 =\ 
& \lambda_{11,11} \left(\Phi_1^\dagger \Phi_1\right)^2 +
\lambda_{22,22} \left(\Phi_2^\dagger \Phi_2\right)^2 +
2 \lambda_{11,22} \left(\Phi_1^\dagger \Phi_1\right)\left(\Phi_2^\dagger \Phi_2\right) +
2 \lambda_{12,21} \left(\Phi_1^\dagger \Phi_2\right)\left(\Phi_2^\dagger \Phi_1\right)
\nonumber\\
& + \left[
\lambda_{12,12} \left(\Phi_1^\dagger \Phi_2\right)^2
+ 2 \lambda_{11,12} \left(\Phi_1^\dagger \Phi_1\right)\left(\Phi_1^\dagger \Phi_2\right)
+ 2 \lambda_{12,22} \left(\Phi_1^\dagger \Phi_2\right)\left(\Phi_2^\dagger \Phi_2\right)
+ \text{h.c.}
\right]
\nonumber\\
& + \alpha_{11,11} \left(\varphi_1^+ \varphi_1^- \right)^2
+ \beta_{11,11} \left(\chi_1^2\right)^2
\nonumber\\
& + \delta_{11,11} \left(\Phi_1^\dagger \Phi_1\right) \left(\varphi_1^+ \varphi_1^-\right)
+ \delta_{12,11} \left(\Phi_1^\dagger \Phi_2\right) \left(\varphi_1^+ \varphi_1^-\right)
+ \delta_{12,11}^* \left(\Phi_2^\dagger \Phi_1\right) \left(\varphi_1^+ \varphi_1^-\right)
+ \delta_{22,11} \left(\Phi_2^\dagger \Phi_2\right) \left(\varphi_1^+ \varphi_1^-\right)
\nonumber\\
& + \gamma_{11,11} \left(\Phi_1^\dagger \Phi_1\right) \left(\chi_1^2\right)
+ \gamma_{12,11} \left(\Phi_1^\dagger \Phi_2\right) \left(\chi_1^2\right)
+ \gamma_{12,11}^* \left(\Phi_2^\dagger \Phi_1\right) \left(\chi_1^2\right)
+ \gamma_{22,11} \left(\Phi_2^\dagger \Phi_2\right) \left(\chi_1^2\right)
+ \zeta_{11,11} \left(\varphi_1^+ \varphi_1^-\right) \left(\chi_1^2\right)
\nonumber\\
& + \left[
\kappa_{12,11} \left(\Phi_1^T \sigma_2 \Phi_2\right) \left(\varphi_1^- \chi_1\right)
- \kappa_{12,11} \left(\Phi_2^T \sigma_2 \Phi_1\right) \left(\varphi_1^- \chi_1\right) + \mathrm{h.c.}
\right],
\end{align}
where we have used the relations in
Eqs.~\eqref{propslambda}, \eqref{propsdelta}, \eqref{propsgamma},
and \eqref{propskappa} to get rid of redundant terms.
This is the simplest model featuring the $\kappa_{ab,cd}$ couplings from Eq.~\eqref{eq:potential}.

\subsubsection{Scattering Matrices}

The set of non-zero scattering matrices are:
\be
M_{|2,2,0\rangle} = 
2 \alpha_{11,11},
\ee
\be
M_{|2, \frac{3}{2}, \frac{1}{2}\rangle} =
\bb
\delta_{11,11} & \delta_{12,11} \\
\delta_{12,11}^* & \delta_{22,11}
\eb,
\ee
\be
M_{|1, \frac{1}{2}, \frac{1}{2}\rangle} =
\left[
\begin{array}{cccc}
 2 \gamma _{11,11} & 2 \gamma
   _{12,11} & 0 & 2 i \kappa
   _{12,11}^* \\
 2 \gamma _{12,11}^* & 2 \gamma
   _{22,11} & -2 i \kappa _{12,11}^*
   & 0 \\
 0 & 2 i \kappa _{12,11} & \delta
   _{11,11} & \delta _{12,11}^* \\
 -2 i \kappa _{12,11} & 0 & \delta
   _{12,11} & \delta _{22,11} \\
\end{array}
\right],
\ee
\be
M_{|1,1,0\rangle} = 
\bb
2 (\lambda_{11,22} - \lambda_{12,21}) & 2 i \sqrt{2} \kappa^*_{12,11} \\
-2 i \sqrt{2} \kappa_{12,11} & 2 \zeta_{11,11}
\eb,
\ee
\be
M_{|2,1,1\rangle} = 
\bb
2 \lambda_{11,11} & 2\sqrt{2} \lambda_{11,12} & 2 \lambda_{12,12} \\
2\sqrt{2} \lambda_{11,12}^* & 2 (\lambda_{11,22} +  \lambda_{12,21}) & 2\sqrt{2} \lambda_{12,22} \\
2 \lambda_{12,12}^* & 2\sqrt{2} \lambda_{12,22}^* & 2 \lambda_{22,22}
\eb,
\ee

\be
M_{|1,0,1\rangle} =
\left[
\begin{array}{cccc}
 2 \lambda _{11,11} & 2 \lambda
   _{11,12}^* & 2 \lambda _{11,12} &
   2 \lambda _{12,21} \\
 2 \lambda _{11,12} & 2 \lambda
   _{11,22} & 2 \lambda _{12,12} & 2
   \lambda _{12,22} \\
 2 \lambda _{11,12}^* & 2 \lambda
   _{12,12}^* & 2 \lambda _{11,22} &
   2 \lambda _{12,22}^* \\
 2 \lambda _{12,21} & 2 \lambda
   _{12,22}^* & 2 \lambda _{12,22} &
   2 \lambda _{22,22} \\
\end{array}
\right],
\ee
\be
M_{|0,0,0\rangle} =
\left[
\begin{array}{cccccc}
 6 \lambda _{11,11} & 6 \lambda
   _{11,12}^* & 6 \lambda _{11,12} &
   2 \left(2 \lambda
   _{11,22}+\lambda _{12,21}\right)
   & \sqrt{2} \delta _{11,11} & 2
   \gamma _{11,11} \\
 6 \lambda _{11,12} & 2
   \left(\lambda _{11,22}+2 \lambda
   _{12,21}\right) & 6 \lambda
   _{12,12} & 6 \lambda _{12,22} &
   \sqrt{2} \delta _{12,11} & 2
   \gamma _{12,11} \\
 6 \lambda _{11,12}^* & 6 \lambda
   _{12,12}^* & 2 \left(\lambda
   _{11,22}+2 \lambda
   _{12,21}\right) & 6 \lambda
   _{12,22}^* & \sqrt{2} \delta
   _{12,11}^* & 2 \gamma _{12,11}^*
   \\
 2 \left(2 \lambda _{11,22}+\lambda
   _{12,21}\right) & 6 \lambda
   _{12,22}^* & 6 \lambda _{12,22} &
   6 \lambda _{22,22} & \sqrt{2}
   \delta _{22,11} & 2 \gamma
   _{22,11} \\
 \sqrt{2} \delta _{11,11} & \sqrt{2}
   \delta _{12,11}^* & \sqrt{2}
   \delta _{12,11} & \sqrt{2} \delta
   _{22,11} & 4 \alpha _{11,11} &
   \sqrt{2} \zeta _{11,11} \\
 2 \gamma _{11,11} & 2 \gamma
   _{12,11}^* & 2 \gamma _{12,11} &
   2 \gamma _{22,11} & \sqrt{2}
   \zeta _{11,11} & 12 \beta
   _{11,11} \\
\end{array}
\right].
\ee

\subsubsection{Perturbative Unitarity Bounds}

Since certain eigenvalues are too complex to evaluate analytically,
the perturbative unitarity constraints are shown only for the remaining eigenvalues
of the zero partial-wave amplitude matrix. These read
\begin{align}
\left|
 \zeta_{11,11} + \lambda_{11,22} - \lambda_{12,21}
 \pm
 \sqrt{
   \left(- \zeta_{11,11} - \lambda_{11,22}
   + \lambda_{12,21} \right)^2
   + 8 |\kappa_{12,11}|^2
 }
\right|
&\leq 8\pi,
\\[2mm]
\frac{1}{2} 
\left|
\delta_{11,11} + \delta_{22,11} 
\pm \sqrt{ \left(\delta_{11,11} - \delta_{22,11} \right)^2 + 4 |\delta_{12,11}|^2} 
\right|
&\leq 8\pi,
\\[2mm]
\left|
2~\alpha_{11,11}
\right|
&\leq 8\pi.
\end{align}

\subsection{$\mathbb{Z}_3-$Symmetric 3HDM}

Now, we consider a model with three $SU(2)$ doublets 
$\Phi_1$, $\Phi_2$, and $\Phi_3$.
Under a $\mathbb{Z}_3$ symmetry, the three doublets transform as
\begin{equation}
\Phi_1 \to \Phi_1,\qquad
\Phi_2 \to e^{i \frac{2\pi}{3}}\Phi_2,\qquad
\Phi_3 \to e^{i \frac{4\pi}{3}}\Phi_3.
\end{equation}
Under this symmetry, the potential becomes
\be
\begin{aligned}
V_4 =\ & \lambda_{11,11} \left(\Phi_1^\dagger \Phi_1\right)^2 + \lambda_{22,22} \left(\Phi_2^\dagger \Phi_2\right)^2 + \lambda_{33,33} \left(\Phi_3^\dagger \Phi_3\right)^2 \\
& + 2 \lambda_{11,22} \left(\Phi_1^\dagger \Phi_1\right) \left(\Phi_2^\dagger \Phi_2\right) + 2 \lambda_{11,33} \left(\Phi_1^\dagger \Phi_1\right) \left(\Phi_3^\dagger \Phi_3 \right) + 2 \lambda_{22,33} \left(\Phi_2^\dagger \Phi_2\right) \left(\Phi_3^\dagger \Phi_3\right) \\
& + 2 \lambda_{12,21} \left(\Phi_1^\dagger \Phi_2 \right)\left(\Phi_2^\dagger \Phi_1\right) + 2 \lambda_{13,31} \left(\Phi_1^\dagger \Phi_3\right)\left(\Phi_3^\dagger \Phi_1 \right) + 2 \lambda_{23,32} \left(\Phi_2^\dagger \Phi_3\right)\left(\Phi_3^\dagger \Phi_2\right) \\
& + 2 \left[ \lambda_{12,13} \left(\Phi_1^\dagger \Phi_2\right)\left(\Phi_1^\dagger \Phi_3\right) + \lambda_{13,23} \left(\Phi_1^\dagger \Phi_3 \right)\left(\Phi_2^\dagger \Phi_3\right) + \lambda_{12,32} \left(\Phi_1^\dagger \Phi_2\right) \left(\Phi_3^\dagger \Phi_2\right) + \textrm{h.c.} \right],
\end{aligned}
\ee
where \(\Phi_1\),\(\Phi_2\) and \(\Phi_3\) denote the three scalar doublets.

\subsubsection{Scattering Matrices}

The scattering matrices are all block diagonal.
We find
\be
M_{|2,1,1\rangle} = 
\textrm{blkdiag}(A,B,C),
\ee
where
\begin{eqnarray}
A
&=&
\bb
2(\lambda_{11,33} + \lambda_{13,31}) & 2\sqrt{2}\lambda_{12,32}\\*[1mm]
2\sqrt{2}\lambda_{12,32}^* & 2\lambda_{22,22} 
\eb,
\\*[1mm]
B
&=&
\bb
2\lambda_{11,11} & 2\sqrt{2}\lambda_{12,13}\\*[1mm]
2\sqrt{2}\lambda_{12,13}^* & 2(\lambda_{22,33} + \lambda_{23,32})
\eb,
\\*[1mm]
C
&=&
\bb
2\lambda_{33,33} & 2\sqrt{2}\lambda_{13,23}^* \\*[1mm]
2\sqrt{2}\lambda_{13,23} & 2(\lambda_{11,22} + \lambda_{12,21})
\eb,
\end{eqnarray}
\be
M_{|1,0,1\rangle} =
\textrm{blkdiag}(D,E,F),
\ee
where
\begin{eqnarray}
D
&=&
\bb
2\lambda_{22,33} & 2\lambda_{13,23}^* & 2\lambda_{12,32} \\
2\lambda_{13,23} & 2\lambda_{11,33} & 2\lambda_{12,13}\\
2\lambda_{12,32}^* & 2\lambda_{12,13}^* & 2\lambda_{11,22}
\eb,
\\*[1mm]
E
&=&
\bb
2\lambda_{11,11} & 2\lambda_{12,21} & 2\lambda_{13,31}\\
2\lambda_{12,21} & 2\lambda_{22,22} & 2\lambda_{23,32}\\
2\lambda_{13,31} & 2\lambda_{23,32} & 2\lambda_{33,33}
\eb,
\\*[1mm]
F
&=&
\bb
2\lambda_{22,33} & 2\lambda_{12,32}^* & 2\lambda_{13,23} \\
2\lambda_{12,32} & 2\lambda_{11,22} & 2\lambda_{12,13} \\
2\lambda_{13,23}^* & 2\lambda_{12,13}^* & 2\lambda_{11,33}
\eb,
\end{eqnarray}

\be
M_{|0,0,0\rangle} =\textrm{blkdiag}(G,H,I),
\ee
where
\begin{eqnarray}
G
&=&
\bb
2(\lambda_{22,33} + 2\lambda_{23,32}) & 6\lambda^*_{13,23} & 6\lambda_{12,32}\\
6\lambda_{13,23} & 2(\lambda_{11,33} + 2\lambda_{13,31}) & 6\lambda_{12,13}\\
6\lambda^*_{12,32} & 6\lambda^*_{12,13} & 2(\lambda_{11,22} + 2\lambda_{12,21})
\eb,
\\*[1mm]
H
&=&
\bb
6\lambda_{11,11} & 2(2\lambda_{11,22} + \lambda_{12,21}) & 2(2\lambda_{11,33} + \lambda_{13,31})\\
2(2\lambda_{11,22} + \lambda_{12,21}) & 6\lambda_{22,22} & 2(2\lambda_{22,33} + \lambda_{23,32})\\
2(2\lambda_{11,33} + \lambda_{13,31}) & 2(2\lambda_{22,33} + \lambda_{23,32}) & 6\lambda_{33,33}
\eb,
\\*[1mm]
I
&=&
\bb
2(\lambda_{22,33} + 2\lambda_{23,32}) & 6\lambda^*_{12,32} & 6\lambda_{13,23} \\
6\lambda_{12,32} & 2(\lambda_{11,22} + 2\lambda_{12,21}) & 6\lambda_{12,13} \\
6\lambda^*_{13,23} & 6\lambda^*_{12,13} & 2(\lambda_{11,33} + 2\lambda_{13,31})
\eb,
\end{eqnarray}
and
\be
M_{|1,1,0\rangle} =
\bb
2(\lambda_{11,22} - \lambda_{12,21}) & 0 & 0 \\
0 & 2(\lambda_{11,33} - \lambda_{13,31}) & 0 \\
0 & 0 & 2(\lambda_{22,33} - \lambda_{23,32})
\eb.
\ee

\subsubsection{Perturbative Unitarity Bounds}

We show the results of partial-wave unitarity bounds to the eigenvalues that can be computed analytically,
\begin{align}
2 |\lambda_{11,22} - \lambda_{12,21}|
&\leq 8\pi,
\\[1mm]
2 |\lambda_{11,33} - \lambda_{13,31}|
&\leq 8\pi,
\\[1mm]
2 |\lambda_{22,33} - \lambda_{23,32}|
&\leq 8\pi,
\\[1mm]
\left|\lambda_{11,11}+\lambda_{22,33}+\lambda_{23,32}\pm\sqrt{8 |\lambda_{12,13}|^2+\left(\lambda_{22,33}+\lambda_{23,32}-\lambda_{11,11}\right){}^2}\right|
&\leq 8\pi,
\\[1mm]
\left|
 \lambda_{22,22} + \lambda_{11,33} + \lambda_{13,31}
\pm \sqrt{ 8 |\lambda_{12,32}|^2 +\left(\lambda_{11,33}+\lambda_{13,31}-\lambda_{22,22}\right){}^2}
\right|
&\leq 8\pi,
\\[1mm]
\left| \lambda_{33,33} +\lambda_{11,22} +  \lambda_{12,21}
\pm \sqrt{8 |\lambda_{13,23}|^2 +\left(\lambda_{11,22}+\lambda_{12,21}-\lambda_{33,33}\right){}^2} \right|
&\leq 8\pi.
\end{align}

\bt
\centering
\caption{Comparison of coupling notation.}
\renewcommand{\arraystretch}{1.3}
\setlength{\tabcolsep}{4pt}
\centering
\begin{tabular}{lccc}
\hline
 & \textbf{Term} & \textbf{Our Notation} & \textbf{Notation of \cite{Bento:2017eti}} \\
\hline\hline
& $(\Phi_1^\dagger \Phi_1)^2$ & $\lambda_{11,11}$ & $r_{1}$ \\
\hline
& $(\Phi_2^\dagger \Phi_2)^2$ & $\lambda_{22,22}$ & $r_{2}$ \\
\hline
& $(\Phi_3^\dagger \Phi_3)^2$ & $\lambda_{33,33}$ & $r_{3}$ \\
\hline
& $(\Phi_1^\dagger \Phi_1)(\Phi_2^\dagger \Phi_2)$ & $2\lambda_{11,22}$ & $2r_{4}$ \\
\hline
& $(\Phi_1^\dagger \Phi_1)(\Phi_3^\dagger \Phi_3)$ & $2\lambda_{11,33}$ & $2r_{5}$ \\
\hline
& $(\Phi_2^\dagger \Phi_2)(\Phi_3^\dagger \Phi_3)$ & $2\lambda_{22,33}$ & $2r_{6}$ \\
\hline
& $(\Phi_1^\dagger \Phi_2)(\Phi_2^\dagger \Phi_1)$ & $2\lambda_{12,21}$ & $2r_{7}$ \\
\hline
& $(\Phi_1^\dagger \Phi_3)(\Phi_3^\dagger \Phi_1)$ & $2\lambda_{13,31}$ & $2r_{8}$ \\
\hline
& $(\Phi_2^\dagger \Phi_3)(\Phi_3^\dagger \Phi_2)$ & $2\lambda_{23,32}$ & $2r_{9}$ \\
\hline
& $(\Phi_1^\dagger \Phi_2)(\Phi_1^\dagger \Phi_3)$ & $2\lambda_{12,13}$ & $2c_{4}$ \\
\hline
& $(\Phi_2^\dagger \Phi_1)(\Phi_3^\dagger \Phi_1)$ & $2\lambda_{12,13}^\ast$ & $2c_{4}^\ast$ \\
\hline
& $(\Phi_1^\dagger \Phi_3)(\Phi_2^\dagger \Phi_3)$ & $2\lambda_{13,23}$ & $2c_{11}$ \\
\hline
& $(\Phi_3^\dagger \Phi_1)(\Phi_3^\dagger \Phi_2)$ & $2\lambda_{13,23}^\ast$ & $2c_{11}^\ast$ \\
\hline
& $(\Phi_1^\dagger \Phi_2)(\Phi_3^\dagger \Phi_2)$ & $2\lambda_{12,32}$ & $2c_{12}$ \\
\hline
& $(\Phi_2^\dagger \Phi_1)(\Phi_2^\dagger \Phi_3)$ & $2\lambda_{12,32}^\ast$ & $2c_{12}^\ast$ \\
\hline
\end{tabular}
\label{notation3hdmZ3}
\et
We compared our results with those presented in \cite{Bento:2017eti},
and found them to be consistent.
Table \ref{notation3hdmZ3} provides a summary of the correspondence
between the quartic couplings used in our study and 
those in \cite{Bento:2017eti}.

\newpage

\section{Overview}
\label{sec:conclusion}

Almost all models addressing the outstanding issues in the SM include extra $SU(2)$
singlet and/or doublet scalars.
In particular, many models addressing the dark matter problem include extra
neutral singlet scalars.
Such models must be subject to theoretical constraints, even before a simulation starts.
Those constraints include boundedness from below, nonexistence of lower lying alternative vacua,
and the perturbative unitarity bounds on $2 \rightarrow 2$ scattering.
In this paper, we address the perturbative partial-wave unitarity for the tree-level
scattering matrix in models with any number of scalar doublets,
neutral singlets, and/or charged singlets.
Enforcing the correct high-energy behavior provides bounds on the quartic couplings,
freeing us from defining the exact nature of the quadratic and cubic couplings.

In contrast, if one wishes to turn such bounds into restrictions on masses, mixing angles
and other directly observable quantities, then one must define the full theory.
That is,
one must define the quadratic and cubic couplings,
a specific vacuum, and the mass matrices must be duly diagonalized.
After this, one would strive to invert the relations,
turning the unitarity bounds on the quartic couplings into
restrictions on combinations of masses and mixing angles.
This is possible in simple models, but increasingly more difficult
as the number of fields increases.
In contrast, although the matrices get larger,
our limits on quartic couplings are always applicable, at least numerically.

We classify the states by the conserved quantum numbers
$Q$, $Y$, and $T$, and show that, once one restricts oneself to the minimal 
set of states providing all inequivalent bounds, the quantum number $Q$ is redundant.
We also discuss examples where the existence of extra symmetries allows for the
inclusion of further quantum numbers in the basis, thus greatly simplifying the
scattering matrices.

We introduce the {\tt Mathematica} notebook {\tt BounDS}
that automatically calculates
the quartic part of the potential \textit{and} the scattering matrices and their
eigenvalues,
for any model with any symmetries (discrete or continuous,
Abelian or non-Abelian).
We present results for a variety of particular models,
and compare with the literature, when available.
Our aim is to help provide complete simulations of models beyond the SM
with a necessary and very powerful tool in parameter restriction.

\section*{Acknowledgments}
\noindent
We are grateful to Rui Santos for reading and commenting on this document.
This work is supported in part by the Portuguese
Fundação para a Ciência e Tecnologia (FCT) through the PRR (Recovery and Resilience
Plan), within the scope of the investment "RE-C06-i06 - Science Plus
Capacity Building", measure "RE-C06-i06.m02 - Reinforcement of
financing for International Partnerships in Science,
Technology and Innovation of the PRR", under the project with
reference 2024.01362.CERN.
The work of the authors is
also supported by FCT under Contracts UIDB/00777/2020, and UIDP/00777/2020.
The FCT projects are partially funded through
POCTI (FEDER), COMPETE, QREN, and the EU.
A.M. was additionally supported by FCT with
PhD Grant No. 2024.01340.BD.

\appendix

\section{Redundant scattering matrices}
\label{app:REDUNDANT}

In any scattering process, conserved quantum numbers constrain the possible initial and final states.
In a globally symmetric $SU(2) \times U(1)$ quantum field theory,
initial states with definite electric charge $Q$ and hypercharge $Y$
can only scatter into states with the same $Q$ and $Y$.
That is because these quantum numbers are protected by the $SU(2) \times U(1)$
symmetry and are, thus, conserved.
For this reason, in building $2 \to 2$ scattering matrices, 
Ref.~\cite{Bento:2017eti} labeled two-particle states by $| Q,Y \rangle$.
For models like the one defined in Section~\ref{sec:themodel}, 
we list in Table.~\ref{tab:basisQY} all possible two-particle states.
\begin{table}[h]
\centering
\caption{Basis of two-particle states labeled by $|Q, Y \rangle$.
This table includes all states, some of which provide redundant information.}
\renewcommand{\arraystretch}{1.3}
\setlength{\tabcolsep}{4pt}
\centering
\begin{tabular}{c | c |  c |  c}
$|Q, Y \rangle$ & State & Conditions & Dimensionality\\
\hline\hline
$|2, 2 \rangle$						
&	$\varphi^+_i \varphi^+_j$	
& $i\leq j$
& $n_c(n_c+1)/2$
\\\hline
$|2, \frac{3}{2}\rangle$	
&	$\phi^+_i \varphi^+_j$		
& 	---
& $n_D n_c$
\\\hline
$|2, 1\rangle$						
&	$\phi^+_i \phi^+_j$		
& $i\leq j$
& $n_D(n_D+1)/2$
\\\hline
$|1, \frac{3}{2}\rangle$						
& $\phi^0_i \varphi^+_j$	
& ---
& $n_D n_c$
\\\hline
$|1, 1\rangle$						
& $\left\{\phi^+_{i}\, \phi^0_{j} , \ \varphi^+_i \chi_j \right\}$	
& ---
&  $n_D^2 + n_n n_c$
\\\hline
$|1, \frac{1}{2}\rangle$	
&	 $\left\{\phi^+_i \chi_j , \ \phi^{0*}_i \varphi^+_j \right\}$
& ---	
&   $n_D (n_n + n_c)$
\\\hline
$|1, 0\rangle$						
&	$\phi^+_i \phi^{0*}_j$	
& ---
&  $n_D^2$
\\\hline
$|0, 1\rangle$						
&	$\phi^0_i \phi^0_j$	
&	$i \leq j$
& $n_D(n_D+1)/2$
\\\hline
$|0, \frac{1}{2}\rangle$	
&	 $\left\{\phi^0_i \chi_j , \ \phi^{-}_i \varphi^+_j \right\}$
&  ---	
&  $n_D (n_n + n_c)$
\\\hline
$|0, 0\rangle$						
& $\left\{ \phi_i^+\phi_j^- , \phi_i^0\phi_j^{0*}
, \  \varphi_i^+ \varphi_j^-,\ \chi_i \chi_j \right\}$
& $\{\, \textrm{---}\,,\, \textrm{---}\,,\, \textrm{---}\,,\, i\leq j \}$	
&  $2 n_D^2 + n_c^2 + n_n(n_n+1)/2$
\\\hline
\end{tabular}
\label{tab:basisQY}
\end{table}
However, not all scattering amplitudes are independent.
Notice that
\begin{eqnarray}
&&\mathcal{M}\left[  \phi^+_a \varphi^+_b \to \phi^+_c \varphi^+_d   \right]
=
\mathcal{M}\left[  \phi^0_a \varphi^+_b \to \phi^0_c \varphi^+_d   \right]
= \delta_{ca,db},\\
&&\mathcal{M}\left[  \phi^+_a \phi^+_b \to \phi^+_c \phi^+_d   \right]
=
\mathcal{M}\left[  \phi^0_a \phi^0_b \to \phi^0_c \phi^0_d   \right]
= 2\left(\lambda_{ca,db}+\lambda_{da,cb}\right),\\
&&\mathcal{M}\left[  \phi^+_a \chi_b \to \phi^+_c \chi_d   \right]
=
\mathcal{M}\left[  \phi^0_a \chi_b \to \phi^0_c \chi_d   \right]
= 2\gamma_{ca,bd},\\
&&\mathcal{M}\left[  \phi^+_a \chi_b \to \phi^{0*}_c \varphi^+_d   \right]
=
\mathcal{M}\left[  \phi^0_a \chi_b \to \phi^{-}_c \varphi^+_d   \right]
= 2i\kappa_{ca,db},\\
&&\mathcal{M}\left[  \phi^{0*}_a \varphi^+_b \to \phi^{0*}_c \varphi^+_d   \right]
=
\mathcal{M}\left[  \phi^{-}_a \varphi^+_b \to \phi^{-}_c \varphi^+_d   \right]
= \delta_{ac,db}.
\end{eqnarray}
Therefore, perturbative unitarity bounds 
obtained from scatterings involving
$|1, \frac{3}{2}\rangle$,
$|0, 1\rangle$	,
and
$|0, \frac{1}{2}\rangle$	
are redundant because they
are identical to those derived from
$|2, \frac{3}{2}\rangle$,
$|2, 1\rangle$	,
and
$|1, \frac{1}{2}\rangle$	, respectively.

In any scattering involving $SU(2)$ doublets, 
total isospin $T$ must also be conserved,
For this reason, two-particle states should be 
further labeled by $|Q, Y, T\rangle$, 
with $T=0,1$ in this class of models.
Using Clebsch-Gordan coefficients, 
we can split the two-particle state $\phi^+_i \phi^0_j$ 
in $|1,1\rangle$ into
\begin{eqnarray}
|1, 1, 0\rangle: &&
\phi^+_{[i}\, \phi^0_{j]} 
\, \equiv \,
\frac{1}{\sqrt{2}}\left(\phi^+_i\, \phi^0_j - \phi^+_j\, \phi^0_i\right),\\
|1, 1, 1\rangle: &&
\phi^+_{(i}\, \phi^0_{j)}
\, \equiv \,
\frac{1}{\sqrt{2}}\left(\phi^+_i\, \phi^0_j + \phi^+_j\, \phi^0_i\right).
\end{eqnarray}
In the same manner, 
we can split the two-particle state $\phi^+_i \phi^-_j$ 
in $|0,0\rangle$ into
\begin{eqnarray}
|0,0, 0\rangle: &&
\Phi_i\, \Phi_j^*
\, \equiv \,
\frac{1}{\sqrt{2}}\left(\phi^+_i\, \phi^-_j + \phi^0_i\, \phi^{0*}_j\right),\\
|0,0, 1\rangle: &&
\overline{\Phi_i\, \Phi_j^*}
\, \equiv \,
\frac{1}{\sqrt{2}}\left(\phi^+_i\, \phi^-_j - \phi^0_i\, \phi^{0*}_j\right).
\end{eqnarray}
These redefinitions of states lead, once again,
to redundant scattering matrices because
\begin{eqnarray}
&&\mathcal{M}\left[  \phi^+_a \phi^+_b \to \phi^+_c \phi^+_d   \right]
=
\mathcal{M}\left[  
\phi^+_{(a}\, \phi^0_{b)}
\to
\phi^+_{(c}\, \phi^0_{d)}   \right]
= 2\left(\lambda_{ca,db}+\lambda_{da,cb}\right),\\
&&\mathcal{M}\left[  \phi^+_a \phi^{0*}_b \to \phi^+_c \phi^{0*}_d   \right]
=
\mathcal{M}\left[  
\overline{\Phi_a \Phi_b^*}
\to
\overline{\Phi_c \Phi_d^*}   \right]
= 2\lambda_{ca,bd}.
\end{eqnarray}
We therefore conclude that it is sufficient
to apply the partial-wave unitarity bounds to 
the scattering matrices built out of the states
listed in Table~\ref{basistable}, \textit{i.e.} the states
\begin{equation}
\left|  Q , Y , T \right\rangle \quad = \quad
\left|  2 , 2 , 0 \right\rangle, \ 
\left|  2 , \frac{3}{2} , \frac{1}{2} \right\rangle,\ 
\left|  2 , 1 , 1 \right\rangle, \ 
\left|  1 , 1 , 0 \right\rangle, \ 
\left|  1 , \frac{1}{2} , \frac{1}{2} \right\rangle, \ 
\left|  1 , 0 , 1 \right\rangle, \ 
\left|  0 , 0 , 0 \right\rangle.
\end{equation}

\section{\label{app:SMexample}The Standard Model example}

\subsubsection*{Labeling states by $Q$ and $Y$}

If we classify the states solely by their electric charge $Q$
and hypercharge $Y$, the SM requires computing the 
scattering matrices for the states listed in Table~\ref{basistablesm}.
\bt
\centering
\renewcommand{\arraystretch}{1.3}
\setlength{\tabcolsep}{4pt}
\centering
\begin{tabular}{c | c | c}
$|Q, Y\rangle$ & State & Dimensionality \\
\hline\hline
$|2, 1\rangle$ & $\phi^+_1 \phi^+_1$ & 1 \\
\hline
$|1, 1\rangle$ & $\phi^+_1 \phi^0_1$ & 1 \\
\hline
$|1, 0\rangle$ & $\phi^+_1 \phi^{0*}_1$ & 1 \\
\hline
$|0, 1\rangle$ & $\phi^0_1 \phi^0_1$ & 1 \\
\hline
$|0, 0\rangle$ & $\left\{\phi^+_1 \phi^-_1,\ \phi^0_1 \phi^{0*}_1 \right\}$ & 2 \\
\hline
\end{tabular}
\caption{Basis of two-particle states labeled by $|Q, Y\rangle$.}
\label{basistablesm}
\et
The corresponding scattering matrices are
\begin{align}
M_{|0,1\rangle} = M_{|1,0\rangle} = M_{|1,1\rangle} = M_{|2,1\rangle} &= 
2\lambda_{11,11}, \\[6pt]
M_{|0,0\rangle} &= 
\bb
4\lambda_{11,11} & 2\lambda_{11,11} \\
2\lambda_{11,11} & 4\lambda_{11,11}
\eb.
\end{align}

\subsubsection*{Including Total Isospin}

We now refine the classification by also incorporating the total isospin $T$.
The two-particulate states with definite $Q$, $Y$, and $T$ are,
thus, the ones listed
in Table~\ref{basistablesmwithT}.
\bt
\centering
\renewcommand{\arraystretch}{1.3}
\setlength{\tabcolsep}{4pt}
\centering
\begin{tabular}{c | c | c}
$|Q, Y, T\rangle$ & State & Dimensionality \\
\hline\hline
$|2, 1, 1\rangle$ & $\phi^+_1 \phi^+_1$ &  1 \\
\hline
$|1, 1, 1\rangle$ & $\phi^+_{(1} \phi^0_{1)}$ & 1 \\
\hline
$|1, 0, 1\rangle$ & $\phi^+_1 \phi^{0*}_1$ & 1 \\
\hline
$|0, 0, 1\rangle$ & $\overline{\Phi_1  \Phi_1^*}$ & 1 \\
\hline
$|0, 0, 0\rangle$ & $\Phi_1 \Phi_1^* $  & 1 \\
\hline
$|0, 1, 1\rangle$ & $\phi^0_1 \phi^0_1$ & 1 \\
\hline
\end{tabular}
\caption{Basis of two-particle states labeled by $|Q, Y, T\rangle$.}
\label{basistablesmwithT}
\et

Let transformation from the old $(Q,Y)$ basis 
to the new $(Q,Y,T)$ basis be given by
\be
\bb
\Phi_1 \Phi_1^* \\
\overline{\Phi_1  \Phi_1^*}
\eb
=
U\,
\bb
\phi_1^+ \phi_1^- \\
\phi_1^0 \phi_1^{0*}
\eb
=
\frac{1}{\sqrt{2}}
\bb
1 & 1 \\
1 & -1
\eb
\bb
\phi_1^+ \phi_1^- \\
\phi_1^0 \phi_1^{0*}
\eb.
\ee
As a result, the scattering matrix for the
states $|0,0\rangle$, 
gets diagonalized as
\be
(M_{|0,0\rangle})_{\text{new}} = U (M_{|0,0\rangle})_{\text{old}} U^\dagger 
= 
\bb
6\lambda_{11,11} & 0 \\
0 & 2\lambda_{11,11}
\eb.
\ee
Similarly, for the $|1,1\rangle$ states,
only the symmetric combination survives
\be
\phi^+_{(1} \phi^0_{1)}
=
\frac{1}{\sqrt{2}}
\bb
1 & 1
\eb
\bb
\phi_1^+ \phi_1^0 \\
\phi_1^0 \phi_1^+
\eb,
\ee
meaning
\be
(M_{|1,1\rangle})_{\text{new}} =
2\lambda_{11,11}.
\ee

Collecting all cases, the scattering eigenvalues are
\begin{align}
M_{|0,0,0\rangle} &= 6\lambda_{11,11}, \\[4pt]
M_{|0,0,1\rangle} &= M_{|1,1,0\rangle} = M_{|2,1\rangle}
= M_{|1,0\rangle} = M_{|0,1\rangle} = 2\lambda_{11,11}.
\end{align}
The eigenvalues, and therefore the unitarity bounds,
follow directly.
This demonstrates the advantage of organizing states in the $|Q,Y,T\rangle$ basis:
the scattering matrices simplify and, in this case, diagonalize naturally.

\section{\label{app:NOTEBOOK}{\tt BounDS}}

We have developed {\tt BounDS}, a \texttt{Mathematica} 
notebook, that automates the process of writing the quartic potential and deriving
partial-wave unitarity bounds for models within the class of models
defined in Section~\ref{sec:themodel}.
Notice that ScannerS \cite{Coimbra:2013qq,Muhlleitner:2020wwk} 
has a tool to calculate the scattering matrices,
but the user must introduce the correct potential.
Here, the user just introduces the number of fields and their symmetries;
our program calculates the correct quartic potential and then the relevant scattering matrices.
As stated, the notebook can be downloaded from
\begin{center}
\linkGIT
\end{center}

The notebook {\tt BounDS} is divided into three parts. 
To illustrate how it works, 
we consider the $U(1)$-symmetric 2HDM. 
In step 1, the user specifies the number of fields to include.
\begin{lstlisting}
nD=2;
nc=0;
nn=0;
\end{lstlisting}
If needed, the user can also impose additional symmetries on the fields
by specifying the number of symmetries {\tt nSym}, and the way
the symmetries act on the fields. The latter is done by
populating the vector {\tt Sym[]}:
\begin{lstlisting}
nSym = 1;

Sym[1] = {
   [CapitalPhi][1] -> [CapitalPhi][1],
   [CapitalPhi][2] -> Exp[I a] [CapitalPhi][2]
   };

Assume = {a [Element] Reals};
\end{lstlisting}
The list {\tt Assume} contains information
about group theoretical parameters, such as
the space they are defined in and relationships
between them.
Furthermore, the function {\tt Conjugate[]}
may be used for symmetries that require conjugation of a field.
{\tt BounDS} has been successfully tested with a wide range of symmetry groups,
including both discrete and continuous, Abelian and non-Abelian cases.
The list {\tt Sym[]} is read sequentially, and some symmetries may take longer to
resolve than others.
For improved time performance,
we recommend that the user declare Abelian symmetries
before imposing possible non-Abelian symmetries.

After evaluating the cells in step 1, 
the user can simply run all the cells in step 2.
Under the hood, this module does the following operations:
\begin{itemize}
	\item Lists the minimal set of linearly independent
quartic couplings allowed by the symmetries.
	\item Assembles the 7 independent scattering matrices
defined in Section~\ref{subsec:ScateringMatrices}.
	\item Block-diagonalizes the scattering matrices by
swapping rows and columns.
\end{itemize}

Finally, step 3 is dedicated to
visualizing and analyzing the output.
By calling {\tt Potential4}, the user can output the
quartic part of the most general scalar potential allowed
by the symmetries of the model. Anywhere in the code,
the native {\tt Mathematica} function {\tt TeXForm} can
be called to get the \LaTeX\ source code:
\begin{lstlisting}
Potential4 //TeXForm
\end{lstlisting}
\begin{eqnarray}
V_4 &=&
\lambda _{11,11} \left(\Phi _1^{\dagger }\Phi _1\right){}^2
+2 \lambda _{11,22} \Phi _2^{\dagger }\Phi _2 \, \Phi _1^{\dagger }\Phi _1+2 \lambda
   _{12,21} \Phi _1^{\dagger }\Phi _2 \,\Phi _2^{\dagger }\Phi _1
+\lambda _{22,22} \left(\Phi _2^{\dagger }\Phi _2\right){}^2.
\end{eqnarray}
Furthermore, the user may access the basis vector 
and the corresponding scattering matrix by 
specifying the value of $Q$, $Y$, and $T$, and evaluating
the functions {\tt Basis[Q, Y, T]} and {\tt ScatteringMatrix[Q, Y, T]},
respectively:
\begin{lstlisting}
Q = 0;
Y = 0;
T = 0;

Basis[Q, Y, T] //TeXForm
ScatteringMatrix[Q, Y, T] //TeXForm
\end{lstlisting}
\begin{eqnarray}
&&\left(
\begin{array}{cccc}
 \Phi _2\Phi _2^{\dagger } ,
 & \Phi _1\Phi _1^{\dagger }, 
 & \Phi _2\Phi _1^{\dagger }, 
 & \Phi _1\Phi _2^{\dagger } \\
\end{array}
\right)\, ,\\
&&\bb
 6 \lambda _{22,22} & 2 \left(2 \lambda _{11,22}+\lambda _{12,21}\right) & 0 & 0 \\
 2 \left(2 \lambda _{11,22}+\lambda _{12,21}\right) & 6 \lambda _{11,11} & 0 & 0 \\
 0 & 0 & 2 \left(\lambda _{11,22}+2 \lambda _{12,21}\right) & 0 \\
 0 & 0 & 0 & 2 \left(\lambda _{11,22}+2 \lambda _{12,21}\right) \\
\eb.
\end{eqnarray}
Note that the ordering of the two-particle states in the 
basis vector may differ from the expected convention. 
This is an artifact of the block-diagonalization routine. 
However, the corresponding eigenvalues are unaffected by this reordering.

Finally, the list of eigenvalues for a given scattering matrix 
can be obtained by calling {\tt EigenList[Q, Y, T]}, 
again, with definite values for $Q$, $Y$, and $T$:
\begin{lstlisting}
Q = 0;
Y = 0;
T = 0;

EigenList[Q, Y, T] //FullSimplify //TeXForm
\end{lstlisting}
\begin{eqnarray}
\begin{array}{l}
 2 \left(\lambda _{11,22}+2 \lambda _{12,21}\right), \\
 2 \left(\lambda _{11,22}+2 \lambda _{12,21}\right), \\
 3 \lambda _{11,11}+3 \lambda _{22,22}-\sqrt{4 \left(2 \lambda _{11,22}
+\lambda _{12,21}\right){}^2+9 \left(\lambda _{11,11}-\lambda
   _{22,22}\right){}^2}\, , \\
 3 \lambda _{11,11}+3 \lambda _{22,22}+\sqrt{4 \left(2 \lambda _{11,22}
+\lambda _{12,21}\right){}^2+9 \left(\lambda _{11,11}-\lambda
   _{22,22}\right){}^2}\, . \\
\end{array}
\end{eqnarray}

\section{$\mathbb{Z}_2-$Symmetric 2HDM}
\label{appD}

In a process involving complex scalars that are singlets under
the SM gauge group, there are CP-even and CP-odd components that scatter independently.
Furthermore, if there are additional flavour symmetries,
each two-particle state must also be labeled by its
corresponding charge $S_\alpha$, following the idea of \cite{Ginzburg:2005dt}.
Therefore, we can label all states by,
\be
    |Q, Y, T, \mathbf{\text{\textbf{CP}}, S_1, S_2, \cdots} \rangle.
    \label{basis}
\ee

Let us consider the $\mathbb{Z}_2$-symmetric 2HDM with
the two-particle states labeled by $|Q, Y, T, \mathbb{Z}_2 \rangle$.
The basis of states is explicitly given in Table~\ref{poiiuytrdf}, where,
in the last two lines, we have used the definition in Eq.~\eqref{def_PhiGrande}.
\bt
\centering
\renewcommand{\arraystretch}{1.3}
\setlength{\tabcolsep}{4pt}
\centering
\begin{tabular}{c | c | c}
$|Q, Y, T, \mathbb{Z}_2\rangle$ & State & Dimensionality \\
\hline\hline
$|2, 1, 1, +1\rangle$ & $\{\phi^+_{1}\phi^+_{1},\,
\phi^+_{2}\phi^+_{2}\}$ &  2 \\
$|2, 1, 1, -1\rangle$ & $\phi^+_{1}\phi^+_{2}$ &  1 \\
$|1, 1, 0, -1\rangle$ & $\phi^+_{[1}\phi^0_{2]}$ &  1 \\
$|1, 0, 1, +1\rangle$ & $\{\phi^+_{1}\phi^{0*}_{1},\,
\phi^+_{2}\phi^{0*}_{2}\}$ &  2 \\
$|1, 0, 1, -1\rangle$ & $\{\phi^+_{1}\phi^{0*}_{2},\,
\phi^+_{2}\phi^{0*}_{1}\}$ &  2 \\
$|0, 0, 0, +1\rangle$ & $\{\Phi_1 \Phi_1^*,\,
\Phi_2 \Phi_2^*\}$ &  2 \\
$|0, 0, 0, -1\rangle$ & $\{\Phi_1 \Phi_2^*,\,
\Phi_2 \Phi_1^*\}$ &  2 \\
\hline
\end{tabular}
\caption{Basis of two-particle states labeled by 
$|Q, Y, T, \mathbb{Z}_2\rangle$.}
\label{poiiuytrdf}
\et

The scattering matrices are thus
\ba
M_{ \left|  2,\, 1,\, 1 ,\, +1 \right\rangle }
&=&
\bb
 2 \lambda _{11,11} & 2 \lambda _{12,12} \\
 2 \lambda _{12,12} & 2 \lambda _{22,22}
\eb,
\\
M_{\left|  2,\, 1,\, 1 ,\, -1 \right\rangle }
&=&
 2 \left(\lambda _{11,22}+\lambda _{12,21}\right),
\\
M_{ \left|  1,\, 1,\, 0 ,\, -1 \right\rangle }
&=&
 2 \left(\lambda _{11,22}-\lambda _{12,21}\right),
\\
M_{ \left|  1,\,0,\, 1 ,\, +1 \right\rangle }
&=&
\bb
 2 \lambda _{11,11} & 2 \lambda _{12,21} \\
 2 \lambda _{12,21} & 2 \lambda _{22,22}
\eb,
\\
M_{ \left|  1,\,0,\, 1 ,\, -1 \right\rangle }
&=&
\bb
 2 \lambda _{11,22} & 2 \lambda _{12,12} \\
 2 \lambda _{12,12} & 2 \lambda _{11,22}
\eb,
\\
M_{ \left|  0,\,0,\, 0 ,\, +1 \right\rangle }
&=&
\bb
 6 \lambda _{11,11} & 2 \left(2 \lambda _{11,22}+\lambda _{12,21}\right) \\
 2 \left(2 \lambda _{11,22}+\lambda _{12,21}\right) & 6 \lambda _{22,22}
\eb,
\\
M_{ \left|  0,\,0,\, 0 ,\, -1 \right\rangle }
&=&
\bb
 2 \left(\lambda _{11,22}+2 \lambda _{12,21}\right) & 6 \lambda _{12,12} \\
 6 \lambda _{12,12} & 2 \left(\lambda _{11,22}+2 \lambda _{12,21}\right)
\eb.
\ea

Using the substitutions in Table~\ref{tab:2HDMnotation}
we recover the well-known unitarity bounds for the $\mathbb{Z}_2$-symmetric 2HDM,
shown in Eqs.~\eqref{Z2-2HDM__1}--\eqref{Z2-2HDM}.

We find that, by labeling states with flavour symmetries in addition to charge,
hypercharge, and total isospin, the scattering matrices simplify further 
and reduce
in dimensionality compared to those in Section~\ref{z22hdm}.
This reduction greatly facilitates the subsequent calculations, as shown in this appendix.

\bibliographystyle{JHEP}
\bibliography{ref}

\end{document}